\documentclass[prb,preprint,showpacs,preprintnumbers,amsmath,amssymb]{revtex4}

\usepackage{graphicx,subfigure}
\usepackage{dcolumn}
\usepackage{bm}
\usepackage{psfrag}
\usepackage{multirow}

\def\be{\begin{equation}}       \def\ee{\end{equation}}
\def\bea{\begin{eqnarray}}      \def\eea{\end{eqnarray}}
\def\half{\frac{1}{2}}
\def\dag{\dagger}
\def\non{\nonumber}
\begin{document}

\title{Second-order quantum corrections for the frustrated, 
spatially anisotropic, spin-1/2 Heisenberg antiferromagnet 
on a square lattice}
\author{Kingshuk Majumdar}
\affiliation{Department of Physics, Grand Valley State University, Allendale, 
Michigan 49401, USA}
\email{majumdak@gvsu.edu}
\date{\today}

\begin{abstract}\label{abstract}
The effects of quantum fluctuations due to directional anisotropy and frustration between nearest neighbors
and next-nearest neighbors of the quantum spin-$1/2$ Heisenberg antiferromagnet on a 
square lattice are investigated using spin-wave expansion. We have calculated the spin-wave energy dispersion in the entire Brillouin zone, renormalized spin-wave velocities, and the magnetization up to second order in $1/S$ expansion for the antiferromagnetic Ne\'{e}l and collinear antiferromagnetic stripe
phases. It is shown that the second-order corrections become significant with increase in frustration. With these corrections magnetizations and spin-wave velocities for both the phases become zero at the quantum critical points as expected from other numerical and analytical methods. We have shown that 
the transition between the two ordered phases are always separated by the disordered paramagnetic phase.
\end{abstract}

\maketitle

\section{\label{sec:Intro}Introduction}

The physics of two-dimensional frustrated spin-1/2 Heisenberg antiferromagnet (HAFM) continues
to attract considerable attention due to the discovery and availability of new magnetic 
materials such as the layered oxide high-temperature superconductors.~\cite{wijn70a,wijn70b,kim99,coldea01,ronnow01,chris04,chris07,bombardi04,melzi00,melzi01,carretta02} These systems can be well described by 
the Heisenberg spin model with nearest neighbor (NN) antiferromagnetic coupling $J_1$ and next-nearest neighbor (NNN) antiferromagnetic coupling $J_2$.  Experimentally by applying high pressures the ground state phase diagram of these frustrated spin systems can be explored from low $\eta=J_2/J_1$ to high $\eta$. For example, Li$_2$VOSiO$_4$ is an insulating vanadium oxide, with spin $s=1/2$ V$^{4+}$ ions arranged in square lattice planes at the centers of VO$_4$ pyramids. These are linked by SiO$_4$ tetrahedra, with Li ions occupying the space between the V-O planes. X-ray diffraction measurements on this compound show that the value of $\eta$ decreases by about 40\% with increase in pressure from zero to 7.6 GPa.~\cite{pavarini08} Moreover, nuclear magnetic resonance, magnetization, specific heat, and 
muon spin rotation measurements on these compounds (Li$_2$VOSiO$_4$, Li$_2$VOGeO$_4$, VOMoO$_4$, BaCdVO(PO$_4$)$_2$) show significant coupling between NN and NNN neighbors.~\cite{melzi00,melzi01,bombardi04} In addition these experiments on Li$_2$VOSiO$_4$ have shown that it undergoes a phase transition at a low temperature (2.8 K) to collinear 
antiferromagnetic order with magnetic moments lying in the $a-b$ plane  with $J_2+J_1 \sim 8.2(1)$ K 
and $J_2/J_1 \sim 1.1 (1)$.~\cite{melzi01,carretta02}

Quantum spin-$1/2$ antiferromagnetic $J_1-J_2$
model on a square lattice has been studied extensively by various analytical
and numerical techniques such as the diagrammatic perturbation 
theory based on spin-wave expansion~\cite{harris71,chandra88,sudip89,castilla91,canali92A,canali92B,igar92,igar93,
igar05,caprioti03,syrom10}, modified spin-wave theory~\cite{dot94}, 
field theory~\cite{valeri99,valeri00,nersesyan03,oleg04,shannon04,isaev09,schmidt2010a},
series expansion~\cite{weihong91,weihong91-2,hamer92,weihong93,oitmaa96,weihong05}, 
exact diagonalization~\cite{sindzingre04}, DMRG~\cite{wang00,hako01,white07}, effective field theory~\cite{viana07,mendonca10}, coupled
cluster method\cite{bishop08}, band-structure calculations,~\cite{tsirlin09}
and Quantum Monte Carlo~\cite{sandvik01,capriotti03,yunoki04}. It is now well known that at low temperatures these systems exhibit new types of magnetic order and novel quantum phases.~\cite{diep,subir1} 
For $J_2=0$ the ground state is antiferromagnetically
ordered at zero temperature. Addition of next nearest neighbor interactions induces
a strong frustration and break the antiferromagnetic (AF) order at $J_2 \sim J_1/2$. The competition between 
NN and NNN interactions for the square lattice is characterized by the frustration parameter 
$\eta= J_2/J_1$. It has been found that a disordered paramagnetic phase exists between 
$\eta_{1c} \approx 0.38$ and $\eta_{2c} \approx 0.60$.~\cite{singh,sushkov}
For $\eta<\eta_{1c}$ the square lattice is AF-ordered whereas for $\eta>\eta_{2c}$ a 
degenerate collinear antiferromagnetic
stripe phase (CAF) emerges. In the collinear state the NN spins have a parallel orientation 
in the vertical direction and antiparallel orientation in the horizontal direction or vice 
versa. The exact nature of the phase transitions and the nature of the intermediate 
phase are still debatable. It is believed that the phase 
transition from the AF-ordered state to the intermediate paramagnetic state at $\eta_{1c}$ is of second order and from the paramagnetic state to the collinear state at $\eta_{2c}$ is of first order.~\cite{singh,sushkov}

A generalization of the frustrated $J_1-J_2$ model is the $J_1-J_1^\prime-J_2$ model where 
$\zeta=J_1^\prime/J_1$ is the directional anisotropy parameter.~\cite{nersesyan03,oleg04} It is known that the spatial anisotropy reduces the width of the disordered phase. Extensive
band structure calculations~\cite{tsirlin09} for the vanadium phosphate compounds Pb$_2$VO(PO$_4$)$_2$, SrZnVO(PO$_4$)$_2$,
BaZnVO(PO$_4$)$_2$, BaCdVO(PO$_4$)$_2$ have shown four different exchange couplings: J$_1$ and J$_1^\prime$ between the NN and J$_2$ and J$_2^\prime$ between NNN. For example $\zeta \approx 0.7$ and $J_2^\prime/J_2 \approx 0.4$ were obtained for SrZnVO(PO$_4$)$_2$. A possible realization of the $J_1-J_1^\prime-J_2$ model may be the compound (NO)Cu(NO$_3$)$_3$~\cite{volkova10} though recent band-structure calculations show a uniform spin chain model with different types of anisotropy and weak interchain couplings~\cite{janson10}. Within the spin-wave expansion the effect of directional anisotropy on the spin-wave energy dispersion
and the transverse dynamical structure factor has been studied before.~\cite{igar05} However, the effect of NNN frustration has not been incorporated in that study. 

For the $J_1-J_1^\prime-J_2$ model using a higher-order coupled cluster method Bishop et al.~\cite{bishop08} reported  existence of a quantum 
triple point (QTP) at $\zeta \approx 0.60,\; \eta \approx 0.33$ . Below this point they predicted a
second-order phase transition between the quantum Ne\'{e}l and stripe phases, whereas above it
these two phases are separated by an intermediate phase. Existence of a  QTP has also
been reported by other authors~\cite{viana07,mendonca10} where they used effective field theory and
effective renormalization group approach to obtain a QTP at $\zeta=0.51,\;\eta \approx0.28$.
In a DMRG study it was predicted that there is no intermediate 
phase (no spin gap) for $\eta$ lower than 0.287 when $\zeta=1$ 
(isotropic case).~\cite{wang00} But more recent DMRG calculations have concluded that a disordered
paramagnetic region persists for all $\eta>0$.~\cite{hako01}

It should be mentioned that the present $J_1-J_1^\prime-J_2$ model was introduced~\cite{nersesyan03} as a two-dimensional (2D) generalization of the frustrated two-leg ladder. However, the phases of the frustrated $J_1-J_1^\prime-J_2$ Heisenberg model differs from the phase diagram of the frustrated spin-1/2 ladder with rung coupling $J_1^\prime$ and diagonal coupling $J_2$.~\cite{liu08,tosh10} In case of the frustrated spin-1/2 ladder for $J_1^\prime <2 J_2$ the ground state is of Haldane type, with two spin-1/2 on the rung forming effective spin-1. On the other hand for $J_1^\prime >2 J_2$ rung pairs form singlets, resulting in the rung-singlet phase.~\cite{oleg04}

Frustrated two leg ladders share some common features with the present 2D $J_1-J_1^\prime-J_2$ model. Using
bosonization calculations it has been shown that a spin gap and dimerization are also present in this case.~\cite{oleg04} However, the presence of this intermediate phase has been questioned and a direct transition from the rung singlet to the Haldane phase has been reported.~\cite{hung06,kim08} Yet evidence of such a dimerized intermediate phase in the two leg model was found numerically in Refs.~[\onlinecite{liu08,tosh10}] up to a certain value of the interchain coupling.

One of the main motivations of this work
is to investigate (within second-order spin-wave expansion) if a disordered paramagnetic region exists for this frustrated, spatially anisotropic $J_1-J_1^\prime-J_2$ model on a square lattice. We find that the intermediate disordered phase exist even for small spatial anisotropies.

In this work we present a comprehensive study of the effect of zero temperature quantum fluctuations on 
the spin wave energy, spin-wave velocities, and magnetization for the two ordered phases of the 
$J_1-J_1^\prime-J_2$ Heisenberg AF on a square lattice. We use spin-wave expansion based on Holstein-Primakoff transformation up to second order to numerically calculate the physical quantities. 
Whenever possible we compare our results with available experimental data on the systems mentioned above and with other existing analytical or numerical results. 
The paper is organized as follows. Section~\ref{sec:model} provides an introduction
to the Hamiltonian for the Heisenberg spin-$1/2$ AF on a spatially 
anisotropic square lattice. The classical ground state
configurations of the model and the different phases are then briefly discussed. In the next
two sections Sec.~\ref{sec:AFphase} and Sec.~\ref{sec:CAFphase} the spin Hamiltonian 
is mapped to the Hamiltonian of interacting spin-wave excitations (magnons) and 
spin-wave expansion up to second order for spin wave energy, spin-wave velocities, and 
staggered magnetizations are presented for the two ordered phases. 
These physical quantities for the two phases are numerically 
calculated and the results are plotted and discussed in Section~\ref{sec:results}. 
Finally we summarize our results in Section~\ref{sec:conclusions}. Appendices~\ref{VertexAF},
~\ref{Greensfunction}, and \ref{VertexCAF} contain details of the formalism.

\section{\label{sec:model} The Model}
We consider a frustrated S=1/2 antiferromagnet with spatial anisotropy on a  $N_L \times N_L$  
square lattice with three types
of exchange interactions between spins: $J_1$ along the $x$ (row) 
directions, 
$J_1^\prime$ along the $y$ (column) directions, and $J_2$ along the diagonals. We assume all interactions to be antiferromagnetic 
and positive i.e. $J_1,J_1^\prime,J_2 >0$. This $J_1-J_1^\prime-J_2$ spin system is 
described by the Heisenberg Hamiltonian
\be
H = \half J_1 \sum_{i=1}^N{\bf S}_{i} \cdot {\bf S}_{i+\delta_x}
+ \half J_1^\prime \sum_{i=1}^N{\bf S}_{i} \cdot {\bf S}_{i+\delta_y} 
+ \half J_2 \sum_{i=1}^N{\bf S}_{i} \cdot {\bf S}_{i+\delta_x + \delta_y},
\label{hamiltonian}
\ee
where $i$ runs over all lattice sites and $i+\delta_x$ ($\delta_x =\pm 1$) 
and $i+\delta_y$ ($\delta_y =\pm 1$) are the 
nearest neighbors to the $i$-th site along the row and the column direction. The third
term represents the interaction between the next-nearest neighbors, which are along the
diagonals. 

\begin{figure}[httb]
\centering
\includegraphics[width=4.0in]{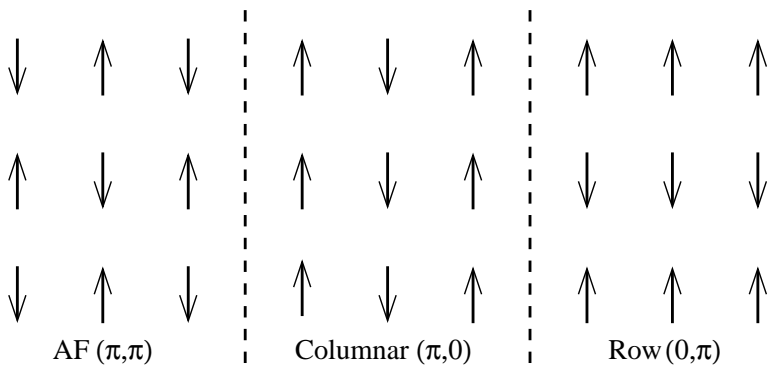}
\caption{\label{fig:GSstates} Classical ground states: (a) AF ($\pi,\pi$), (b) Columnar
($\pi,0$), and (c) Row ($0,\pi$).}
\end{figure}

At zero temperature this model exhibits three types of classical ground state 
configurations: the Ne\'{e}l state or the ($\pi,\pi$) state and the two stripe states which are the 
columnar stripe ($\pi,0$) and the row stripe ($0,\pi$). The spin orientations of these three
states are shown in Fig.~\ref{fig:GSstates}. The Ne\'{e}l state breaks the SU(2) 
and the lattice translational symmetry, but preserves the 
fourfold rotational symmetry $C_4$ of the square. The stripe states 
break SU(2) and partial lattice translational symmetries (along
one direction). In addition this state breaks the invariance under $\pi/2$ real-space 
rotations $C_4$ to $C_2$. 

The classical ground state energies of these states are determined by treating the spins
as classical vectors and then minimizing the energy. These are
\bea
E^{cl}_{\rm AF}/N &=& -\half J_1S^2 z \left[1+\zeta - 2 \eta\right], \nonumber \\
E^{cl}_{\rm columnar}/N &=& -\half J_1S^2 z \left[1-\zeta + 2 \eta\right],\label{cgs}\\
E^{cl}_{\rm row}/N &=& -\half J_1S^2 z \left[-1+\zeta + 2 \eta\right].\nonumber
\eea
Here $\zeta = J_1^\prime/J_1$ is the directional
anisotropy parameter and $\eta=J_2/J_1$ is the magnetic frustration between the 
NN (row direction) and NNN spins. $z=2$ is the number of nearest neighbor sites. 
Eq.~\ref{cgs} shows that the classical ground state is either the antiferromagnetic
Ne\'{e}l (AF) state for $\eta<\zeta/2$ or the columnar antiferromagnetic stripe state (CAF) if 
$\eta>\zeta/2$. The classical first-order phase transition between the AF and CAF state
occurs at the critical value $\eta_c^{\rm class}= \zeta/2$.~\cite{bishop08}

At low  temperature quantum fluctuations play a 
significant role on the phase diagram of the system. In the next 
sections we will consider the classical spins as quantum spins and study the role of 
quantum fluctuations on the AF and CAF ordered phases. We follow a standard procedure by
first expressing the fluctuations around the ``classical" ground state in terms of the
boson operators using the Holstein-Primakoff transformation.~\cite{holstein} The quadratic term in boson
operators corresponds to the linear spin-wave theory, whereas the higher-order terms 
represent spin-wave (magnon) interactions. We keep terms up to second order in 1/S. In the
next step we calculate the renormalized magnon Green's functions and self-energies. Finally,
we calculate the magnon energy dispersion, renormalized 
spin-wave velocities, and the staggered magnetization per spin to the leading order in 1/S$^2$ for the AF and CAF phases. 

\subsection{\label{sec:AFphase}AF Phase - Formalism}
For the AF ordered phase NN interactions are between A and B sublattices and 
NNN interactions are between A-A and B-B sublattices. The Hamiltonian in Eq.~\ref{hamiltonian}
takes the form:
\be
H = J_1 \sum_{i}{\bf S}_{i}^{\rm A} \cdot {\bf S}_{i+\delta_x}^{\rm B}
+ J_1^\prime \sum_{i}{\bf S}_{i}^{\rm A} \cdot {\bf S}_{i+\delta_y}^{\rm B} 
+ \half J_2 \sum_{i}\Big[ {\bf S}_{i}^{\rm A} \cdot {\bf S}_{i+\delta_x + \delta_y}^{\rm A}
+ {\bf S}_{i}^{\rm B} \cdot {\bf S}_{i+\delta_x + \delta_y}^{\rm B}\Big].
\label{ham-AF}
\ee
This Hamiltonian can  be mapped into an equivalent Hamiltonian
of interacting bosons by transforming the spin operators to bosonic creation 
and annihilation operators 
$a^\dag, a$ for ``up''  and $b^\dag, b$ for ``down'' sublattices using the 
Holstein-Primakoff transformations keeping only 
terms up to the order of $1/S^2$ 
\begin{eqnarray}
S_{Ai}^+ &\approx& \sqrt{2S}\Big[1- \half \frac {a_i^\dag a_i}{(2S)}
-\frac 1{8} \frac{a_i^\dag a_i a_i^\dag a_i}{(2S)^2} \Big]a_i,\non \\
S_{Ai}^- &\approx& \sqrt{2S}a_i^\dag \Big[1-\half \frac {a_i^\dag a_i}{(2S)} 
-\frac 1{8} \frac{a_i^\dag a_i a_i^\dag a_i}{(2S)^2} \Big], \non \\
S_{Ai}^z &=& S-a^\dag_ia_i,  \label{holstein}  \\ 
S_{Bj}^+ &\approx& \sqrt{2S}b_j^\dag \Big[1-\half \frac {b_j^\dag b_j}{(2S)} 
-\frac 1{8} \frac{b_j^\dag b_j b_j^\dag b_j}{(2S)^2}\Big],\non \\
S_{Bj}^- &\approx& \sqrt{2S}\Big[1-\half \frac {b_j^\dag b_j}{(2S)}
 -\frac 1{8} \frac{b_j^\dag b_j b_j^\dag b_j}{(2S)^2} \Big]b_j, \non \\
S_{Bj}^z &=& -S+b^\dag_jb_j. \non 
\end{eqnarray}
Substituting Eqs.~\ref{holstein} we expand the Hamiltonian in powers of 
1/S as
\be
H = -\half N J_1 S^2 z(1+\zeta)\Big[1 - \frac {2\eta}{1+\zeta}\Big] + H_0 + H_1+H_2 + ....
\ee 
The first term corresponds to the classical energy of the AF ground state (Eq.~\ref{cgs}).
Next using the spatial Fourier transforms
\[
a_i = \sqrt{\frac 2{N}}\sum_{\bf k} e^{-i{\bf k \cdot R_i}}a_{\bf k},\;\;\;\;
b_j = \sqrt{\frac 2{N}}\sum_{\bf k} e^{-i{\bf k \cdot R_j}}b_{\bf k},
\]
the real space Hamiltonian is transformed to the ${\bf k}$-space Hamiltonian. Momentum
${\bf k}$ is defined in the first Brillouin zone (BZ): $-\pi < k_x \leq \pi,\; 
-\pi <k_y \leq \pi$ (with unit lattice spacing). 
The reduced Brillouin zone contains $N/2$ ${\bf k}$ vectors as the unit cell is 
a magnetic supercell consisting of an $A$-site and a $B$-site.

Furthermore, we diagonalize the quadratic part $H_0$ by transforming the 
operators $a_{\bf k}$ and $b_{\bf k}$ to magnon operators 
$\alpha_{\bf k}$ and $\beta_{\bf k}$ using the  Bogoliubov (BG) transformations
\be
a^\dag_{\bf k} =l_{\bf k} \alpha_{\bf k}^\dag + m_{\bf k}\beta_{-{\bf k}},\;\;\;
b_{-\bf k} =m_{\bf k} \alpha_{\bf k}^\dag + l_{\bf k}\beta_{-{\bf k}},
\ee
where the coefficients $l_{\bf k}$ and $m_{\bf k}$ are defined as
\be
l_{\bf k} = \Big[\frac {1+\epsilon_{\bf k}}{2\epsilon_{\bf k}} \Big]^{1/2},\;\;
m_{\bf k} = -{\rm sgn}(\gamma_{\bf k})\Big[\frac {1-\epsilon_{\bf k}}
{2\epsilon_{\bf k}} \Big]^{1/2}\equiv -x_{\bf k}l_{\bf k},
\ee
with 
\bea
\epsilon_{\bf k} &=& (1-\gamma_{\bf k}^2)^{1/2}, \non \\
\gamma_{\bf k} &=& \gamma_{1{\bf k}}/\kappa_{\bf k}, \non \\
\gamma_{1{\bf k}}&=&[\cos (k_x)+\zeta \cos (k_y)]/(1+\zeta), \label{defs-AF} \\
\gamma_{2{\bf k}} &=& \cos (k_x)\cos(k_y), \non \\
\kappa_{\bf k} &=& 1- \frac {2\eta}{1+\zeta} (1-\gamma_{2{\bf k}}).\non
\eea
$\gamma_{\bf k}$ is 
negative in certain parts of the first BZ - so it is essential to keep track of the sign
of $\gamma_{\bf k}$ through the function ${\rm sgn} (\gamma_{\bf k})$.
After these transformations, the quadratic part of the Hamiltonian becomes
\be
H_0 = J_1Sz(1+\zeta)\sum_{\bf k} \kappa_{\bf k}\left(\epsilon_{\bf k} -1\right)
+J_1Sz(1+\zeta)\sum_{\bf k}\kappa_{\bf k} \epsilon_{\bf k}
\left( \alpha^\dag_{\bf k}\alpha_{\bf k}+\beta^\dag_{\bf k}\beta_{\bf k}\right).
\label{H0term}
\ee
The first term is the zero-point energy and the second term represents the 
excitation energy of the magnons within linear spin-wave theory (LSWT).

The part $H_1$ corresponds to $1/S$ correction to the Hamiltonian. We follow the same
procedure as described above. The resulting expression after
transforming the bosonic operators to the magnon operators is
\bea
H_1 &=& \frac {J_1Sz(1+\zeta)}{2S}\sum_{\bf k}
\Big[ A_{\bf k}\left(\alpha^\dag_{\bf k}\alpha_{\bf k}+
\beta^\dag_{\bf k}\beta_{\bf k}\right) 
+ B_{\bf k}\left(\alpha^\dag_{\bf k}\beta_{-\bf k}+
\beta_{-\bf k}\alpha_{\bf k}\right)\Big] \non \\
&-& \frac {J_1Sz(1+\zeta)}{2SN}\sum_{1234}
\delta_{\bf G}(1+2-3-4)l_1l_2l_3l_4\Big[\alpha_1^\dag \alpha_2^\dag \alpha_3 \alpha_4
V_{1234}^{(1)} +\beta^\dag_{-3}\beta^\dag_{-4}\beta_{-1}\beta_{-2}V_{1234}^{(2)} \non \\
&+&4\alpha_1^\dag \beta_{-4}^\dag \beta_{-2}\alpha_3 V_{1234}^{(3)} +\Big\{
2\alpha_1^\dag \beta_{-2}\alpha_3 \alpha_4V_{1234}^{(4)} +2\beta_{-4}^\dag \beta_{-1}
\beta_{-2}\alpha_3 V_{1234}^{(5)} + \alpha_1^\dag \alpha_2^\dag \beta_{-3}^\dag 
\beta_{-4}^\dag V_{1234}^{(6)} \non \\
&+& h.c.\Big\}\Big].
\label{H1term}
\eea 
In the above equation momenta ${\bf k}_1, {\bf k}_2, {\bf k}_3, {\bf k}_4$ are abbreviated
as 1, 2, 3, and 4. The first term in Eq.~\ref{H1term}, which is known as 
the Oguchi correction~\cite{oguchi} in the literature is obtained by setting the products
of four boson operators into normal ordered forms with respect to the magnon
operators, where $A_{\bf k}$ and $B_{\bf k}$ are 
\bea
A_{\bf k}&=& A_1 \frac 1{\kappa_{\bf k}\epsilon_{\bf k}}\Big[\kappa_{\bf k}
-\gamma_{1{\bf k}}^2\Big] + A_2 \frac 1{\epsilon_{\bf k}}
\Big[1-\gamma_{2{\bf k}}\Big], \\
B_{\bf k} &=& B_1 \frac {1}{\kappa_{\bf k}\epsilon_{\bf k}}
\gamma_{1{\bf k}}\Big[1-\gamma_{2{\bf k}}\Big],
\eea
with
\bea
A_1 &=& \frac 2{N} \sum_{\bf p} \frac 1{\epsilon_{\bf p}}
\Big[\frac {\gamma_{1{\bf p}}^2}{\kappa_{\bf p}}+\epsilon_{\bf p}-1\Big], \\
A_2 &=& \Big(\frac {2\eta}{1+\zeta} \Big)\frac 2{N} \sum_{\bf p} 
\frac 1{\epsilon_{\bf p}}\Big[1-\epsilon_{\bf p}-\gamma_{2{\bf p}}\Big], \\
B_1 &=& \Big(\frac {2\eta}{1+\zeta} \Big)\frac 2{N} \sum_{\bf p} 
\frac 1{\epsilon_{\bf p}}\Big[\gamma_{2{\bf p}}-
\frac {\gamma_{1{\bf p}}^2}{\kappa_{\bf p}}\Big].
\eea
The second term in Eq.~\ref{H1term} represents scattering between spin-waves where the delta
function $\delta_{\bf G}(1+2-3-4)$ ensures that momentum is conserved within a 
reciprocal lattice vector ${\bf G}$. Explicit
forms of the vertex factors $V_{1234}^{i=1...6}$ are given in Appendix~\ref{VertexAF}.

The second order term, $H_2$ is composed of six boson operators. Before the BG transformation
$H_2$ is of the following form:
\bea
H_2&=& \frac {J_1Sz(1+\zeta)}{(2S)^2 N^2}\sum_{123456} \delta_{\bf G}(1+2+3-4-5-6)
\Big[\gamma_1(2+3-6)a_1^\dag a_4 a_5 b^\dag_{-6}b_{-2}b_{-3} \non \\
&+& \gamma_1(3-5-6)a_1^\dag a^\dag_2 a_4 b^\dag_{-5}b_{-6}^\dag b_{-3}  
-\half \Big\{\gamma_1(4)a_4b^\dag_{-5}b_{-1}b_{-6}^\dag b_{-2}b_{-3}
+\gamma_1(3)a_1^\dag a_4 a^\dag_{2}a_5 a_6 b_{-3} + h.c.\Big\} \non \\
&+&  \Big(\frac {2\eta}{1+\zeta}\Big)\Big\{\gamma_2(2+3-6)a_1^\dag a_4a_5a_2^\dag a_3^\dag a_6
+\gamma_2(3-5-6)a_1^\dag a_2^\dag a_4a_3^\dag a_5 a_6 \non \\
&-& \half \Big(
\gamma_2(3)a_1^\dag a_4 a_2^\dag a_5 a_6 a_3^\dag +\gamma_2(1)a_1^\dag a_2^\dag a_4a_3^\dag
a_5 a_6 + h.c. \Big) + a \leftrightarrow b\Big\}\Big].
\eea
After transformation to magnon operators 
$\alpha_{\bf k},\beta_{\bf k}$ the Hamiltonian
in normal ordered form reduces to
\be
H_2= \frac {J_1Sz(1+\zeta)}{(2S)^2} \sum_{\bf k} 
\Big[ C_{1{\bf k}}\left(\alpha^\dag_{\bf k}\alpha_{\bf k}+\beta^\dag_{\bf k}\beta_{\bf k}
\right)+C_{2{\bf k}}\left(\alpha^\dag_{\bf k}\beta_{-\bf k}^\dag+
\beta_{-\bf k}\alpha_{\bf k}\right)+...\Big].
\label{H2term}
\ee
The dotted terms contribute to higher than second order corrections and are thus omitted in 
our calculations.  
The coefficients $C_{1{\bf k}}$ and $C_{2{\bf k}}$ are given in Appendix~\ref{Greensfunction}.
We will find that these corrections play a significant role in the magnon energy dispersion and in the phase diagram for large frustration and/or small anisotropy.

The quasiparticle energy ${\tilde E_{\bf k}^{\rm AF}}$ for magnon excitations, measured 
in units of $J_1Sz(1+\zeta)$ up to second order in $1/S$ is given as
\be
{\tilde E_{\bf k}^{\rm AF}} = E_{\bf k} + \frac 1{(2S)} A_{\bf k}+
\frac 1{(2S)^2}\Big[\Sigma^{(2)}_{\alpha \alpha}({\bf k},E_{\bf k})-
\frac {B_{\bf k}^2}{2E_{\bf k}} \Big].
\label{energyEk-AF}
\ee
Expressions for the magnon Green's functions and self-energies are given in Appendix~\ref{Greensfunction}.

We now define the renormalized spin-wave velocities along the $x$ and $y$ directions at the zone 
boundary using Eq.~\ref{energyEk-AF} as
$V_x=\lim_{k_x \rightarrow 0} 2J_1S(1+\zeta){\tilde E_{\bf k}^{\rm AF}}/k_x$ with $k_y=0$
and $V_y=\lim_{k_y \rightarrow 0} 2J_1S(1+\zeta){\tilde E_{\bf k}^{\rm AF}}/k_y$ with $k_x=0$.
The renormalization factors are expressed as,
\bea
Z_{v_x}^{\rm AF} &\equiv& \frac {V_x}{2J_1S\sqrt{1+\zeta}} = v_{0x}+\frac {v_{1x}}{(2S)}+
\frac {v_{2x}}{(2S)^2}, \\
Z_{v_y}^{\rm AF} &\equiv& \frac {V_y}{2J_1S\sqrt{1+\zeta}} = v_{0y}+\frac {v_{1y}}{(2S)}+
\frac {v_{2y}}{(2S)^2},
\label{vel-AF}
\eea
where 
\bea
v_{0x} &=& (1-2\eta)^{1/2}, \\
v_{1x} &=&(1-2\eta)^{-1/2}\Big[(1-\eta)A_1+\half (1+\zeta)A_2\Big], \\
v_{2x} &=& (1+\zeta)^{1/2} \lim_{k_x \rightarrow 0} \frac 1{k_x} 
\Big[\Sigma^{(2)}_{\alpha \alpha}({\bf k},E_{\bf k})-
\frac {B_{\bf k}^2}{2E_{\bf k}} \Big], \\
v_{0y} &=& (\zeta-2\eta)^{1/2}, \\
v_{1y} &=&(\zeta-2\eta)^{-1/2}\Big[(\zeta-\eta)A_1+\half (1+\zeta)A_2\Big], \\
v_{2y} &=& (1+\zeta)^{1/2} 
\lim_{k_y \rightarrow 0} \frac 1{k_y} \Big[\Sigma^{(2)}_{\alpha \alpha}({\bf k},E_{\bf k})-
\frac {B_{\bf k}^2}{2E_{\bf k}} \Big]. 
\eea
The magnetization $M$ defined as the average of the spin operator
$S_z$ on a given sublattice (say A) is expressed as
\be
M = S-\langle a^\dag_i a_i \rangle = S-\Delta S + \frac {M_1}{(2S)}+\frac {M_2}{(2S)^2},
\label{Mag-AF}
\ee
where
\bea
\Delta S &=& \frac 1{N} \sum_{\bf k} \Big(\frac 1{\epsilon_{\bf k}}-1 \Big), 
\label{MagAF-LSWT} \\
M_1 &=& \frac 2{N} \sum_{\bf k}\frac {l_{\bf k}m_{\bf k}B_{\bf k}}{E_{\bf k}},  
\label{M1-AF} \\
M_2 &=& \frac 2{N} \sum_{\bf k} \Big\{ -(l_{\bf k}^2+m_{\bf k}^2)\frac {B_{\bf k}^2}{4E_{\bf k}^2}
+ \frac {l_{\bf k}m_{\bf k}}{E_{\bf k}}\Sigma^{(2)}_{\alpha \beta}({\bf k},-E_{\bf k}) \non \\
&-& \Big(\frac 2{N} \Big)^2 \sum_{\bf pq} 2l_{\bf k}^2l_{\bf p}^2l_{\bf q}^2l_{\bf k+p-q}^2 
\Big[\frac {(l_{\bf k}^2+m_{\bf k}^2)|V^{(6)}_{\bf k,p,q,[k+p-q]}|^2}
{(E_{\bf k}+E_{\bf p}+E_{\bf q}+E_{\bf k+p-q})^2} \non \\
&+& \frac {2l_{\bf k}m_{\bf k}{\rm sgn}(\gamma_{\bf G})V^{(4)}_{\bf k,p,q,[k+p-q]}
V^{(6)}_{\bf k,p,q,[k+p-q]}}{E_{\bf k}^2-(E_{\bf p}+E_{\bf q}+E_{\bf k+p-q})^2}. \Big]
\label{M2-AF}
\Big\}
\eea
The zeroth-order term $\Delta S$ corresponds to the reduction of magnetization within 
LSWT, $M_1$ term corresponds to the first-order 1/S correction, and $M_2$ is the second-order
correction.

\subsection{\label{sec:CAFphase}CAF Phase - Formalism}
\subsubsection{\label{sec: model-CAF}Hamiltonian}
In the CAF phase ``up'' and ``down'' spins interact along the row directions (NN coupling) and also
along the diagonals (NNN coupling) whereas ``up''-``up'' and ``down''-``down'' spins interact along
the column direction (NN coupling). The Hamiltonian for this phase is described by
\be
H = J_1 \sum_{i}{\bf S}_{i}^{\rm A} \cdot {\bf S}_{i+\delta_x}^{\rm B}
+ \half J_1^\prime\sum_{i}\Big[ {\bf S}_{i}^{\rm A} \cdot {\bf S}_{i+\delta_x + \delta_y}^{\rm A}
+ {\bf S}_{i}^{\rm B} \cdot {\bf S}_{i+\delta_x + \delta_y}^{\rm B}\Big]
+ J_2 \sum_{i}{\bf S}_{i}^{\rm A} \cdot {\bf S}_{i+\delta_y}^{\rm B} .
\label{ham-CAF}
\ee
The Hamiltonians for the AF and the CAF ordered phases (Eq.~\ref{ham-AF} and Eq.~\ref{ham-CAF})
show the similarity between these two phases.
In the AF-phase $J_2$ interactions play the role of $J_1^\prime$ interactions in the CAF
phase. For the CAF-phase the structure factors $\gamma_{1{\bf k}}^\prime,\; 
\gamma_{2{\bf k}}^\prime$ along with other quantities required for the calculations are
defined as
\bea
\gamma^\prime_{1{\bf k}}&=&\big[\cos (k_x)(1+2\eta \cos (k_y))\big]/(1+2\eta), \non\\
\gamma^\prime_{2{\bf k}} &=& \cos(k_y), \non\\
\gamma^\prime_{\bf k} &=& \gamma^\prime_{1{\bf k}}/\kappa^\prime_{\bf k},\\
\kappa_{\bf k}^\prime &=& 1- \frac {\zeta}{1+2\eta} (1-\gamma^\prime_{2{\bf k}}),\non\\
\epsilon_{\bf k}^\prime &=& [1-\gamma_{\bf k}^{\prime 2}]^{1/2}. \non  
\eea
The coefficients for the Oguchi correction that appear in the Hamiltonian $H_1$ are 
\bea
A_{\bf k}^\prime &=& A_1^\prime \frac 1{\kappa^\prime_{\bf k}\epsilon^\prime_{\bf k}}
\Big[\kappa^\prime_{\bf k}
-\gamma_{1{\bf k}}^{\prime 2}\Big] + A_2^\prime \frac 1{\epsilon^\prime_{\bf k}}
\Big[1-\gamma^\prime_{2 {\bf k}}\Big], \\
B_{\bf k}^\prime &=& B_1^\prime \frac {1}{\kappa^\prime_{\bf k}\epsilon^\prime_{\bf k}}
\gamma^\prime_{1 {\bf k}}\Big[1-\gamma^\prime_{2{\bf k}}\Big],
\label{definitions-CAF} 
\eea
with
\bea
A_1^\prime &=& \frac 2{N} \sum_{\bf p} \frac 1{\epsilon^\prime_{\bf p}}
\Big[\frac {\gamma_{1{\bf p}}^{\prime 2}}{\kappa^\prime_{\bf p}}+\epsilon^\prime_{\bf p}-1\Big], \\
A_2^\prime &=& \Big(\frac {\zeta}{1+2\eta} \Big)\frac 2{N} \sum_{\bf p} 
\frac 1{\epsilon^\prime_{\bf p}}\Big[1-\epsilon^\prime_{\bf p}-\gamma^\prime_{2{\bf p}}\Big], \\
B_1^\prime &=& \Big(\frac {\zeta}{1+2\eta} \Big)\frac 2{N} \sum_{\bf p} 
\frac 1{\epsilon^\prime_{\bf p}}\Big[\gamma^\prime_{2{\bf p}}-
\frac {\gamma_{1{\bf p}}^{\prime 2}}{\kappa^\prime_{\bf p}}\Big].
\eea
$H_0,\;H_1$, and $H_2$ can be expressed in the same forms as in Eqs.~\ref{H0term},~\ref{H1term},
and ~\ref{H2term} with the new coefficients $A_{\bf k}^\prime, B_{\bf k}^\prime,C_{1{\bf k}}^\prime, 
C_{2{\bf k}}^\prime$ and with the replacement $\zeta \leftrightarrow 2\eta$. 
The expressions for the two vertex factors $V^{\prime (4)}, V^{\prime (6)}$ and the 
coefficients $C_{1{\bf k}}^\prime, C_{2{\bf k}}^\prime $ are given 
in Appendix ~\ref{VertexCAF}. 
As an example for the CAF phase Eq.~\ref{H0term} takes the form: 
\be
H_0 = J_1Sz(1+2\eta)\sum_{\bf k} \kappa^\prime_{\bf k}\left(\epsilon^\prime_{\bf k} -1\right)
+J_1Sz(1+2\eta)\sum_{\bf k}\kappa^\prime_{\bf k} \epsilon^\prime_{\bf k}
\left( \alpha^\dag_{\bf k}\alpha_{\bf k}+\beta^\dag_{\bf k}\beta_{\bf k}\right),
\label{H0term-CAF}
\ee
The quasiparticle energy ${\tilde E_{\bf k}^{\rm CAF}}$ for magnon excitations, measured 
in units of $J_1Sz(1+2\eta)$ up to second order 
in $1/S$ is given as
\be
{\tilde E_{\bf k}^{\rm CAF}} = E_{\bf k}^\prime + \frac 1{2S} A_{\bf k}^\prime+
\frac 1{(2S)^2}\Big[\Sigma^{\prime (2)}_{\alpha \alpha}({\bf k},E_{\bf k}^\prime)-
\frac {B_{\bf k}^{\prime 2}}{2E_{\bf k}^\prime} \Big]
\label{energyEk-CAF}
\ee

The renormalized spin-wave velocities along the $x$ and $y$ 
directions for this phase are defined as
$V_x=\lim_{k_x \rightarrow 0} 2J_1S(1+2\eta){\tilde E_{\bf k}^{\rm CAF}}/k_x$ with $k_y=0$
and $V_y=\lim_{k_y \rightarrow 0} 2J_1S(1+2\eta){\tilde E_{\bf k}^{\rm CAF}}/k_y$ with $k_x=0$.
The renormalization factors are,
\bea
Z^{\rm CAF}_{v_x} &\equiv& \frac {V^\prime_x}{2J_1S(1+2\eta)} = v^\prime_{0x}+\frac {v^\prime_{1x}}{(2S)}+
\frac {v^\prime_{2x}}{(2S)^2}, \\
Z^{\rm CAF}_{v_y} &\equiv& \frac {V^\prime_y}{2J_1S(1+2\eta)} = v^\prime_{0y}+\frac {v^\prime_{1y}}{(2S)}+
\frac {v^\prime_{2y}}{(2S)^2},
\label{vel-CAF}
\eea
where 
\bea
v^\prime_{0x} &=& 1, \\
v^\prime_{1x} &=& A_1^\prime, \\
v^\prime_{2x} &=& \lim_{k_x \rightarrow 0}\frac 1{k_x} \Big[\Sigma^{\prime (2)}_{\alpha \alpha}({\bf k},E^\prime_{\bf k})-
\frac {B_{\bf k}^{\prime 2}}{2E^\prime_{\bf k}} \Big], \\
v^\prime_{0y} &=& (2\eta+1)^{-1/2}(2\eta-\zeta)^{1/2}, \\
v^\prime_{1y} &=&(2\eta+1)^{-1/2}(2\eta-\zeta)^{-1/2} 
\Big[ (2\eta-\frac {\zeta}{2})A_1+ \half (2\eta+1)A_2 \Big], \\
v^\prime_{2y} &=& \lim_{k_y \rightarrow 0} \frac 1{k_y}\Big[\Sigma^{\prime (2)}_{\alpha \alpha}({\bf k},E^\prime_{\bf k})-
\frac {B_{\bf k}^{\prime 2}}{2E^\prime_{\bf k}} \Big], 
\eea
\section{\label{sec:results}Results}
\subsection{\label{sec: AFphase-result}AF Phase}
\subsubsection{\label{sec: AFphase-energy}Spin-wave energy dispersion}
We numerically evaluate Eq.~\ref{energyEk-AF} to obtain the spin-wave energy 
$2J_1S(1+\zeta){\tilde E_{\bf k}^{\rm AF}}$ as a function of momentum for several values
of $\zeta$ and $\eta$. For the numerical summation, the first BZ is divided into 
$N_L^2$ meshes with $N_L=64$ and then $4096$ points of ${\bf p}$ and 
$4096$ points of ${\bf q}$ are summed up to evaluate the third term in Eq.~\ref{energyEk-AF}. For some 
of the cases we have used $N_L=96$ for better accuracies.

\begin{figure}[httb]
\centering
\includegraphics[width=3.5in]{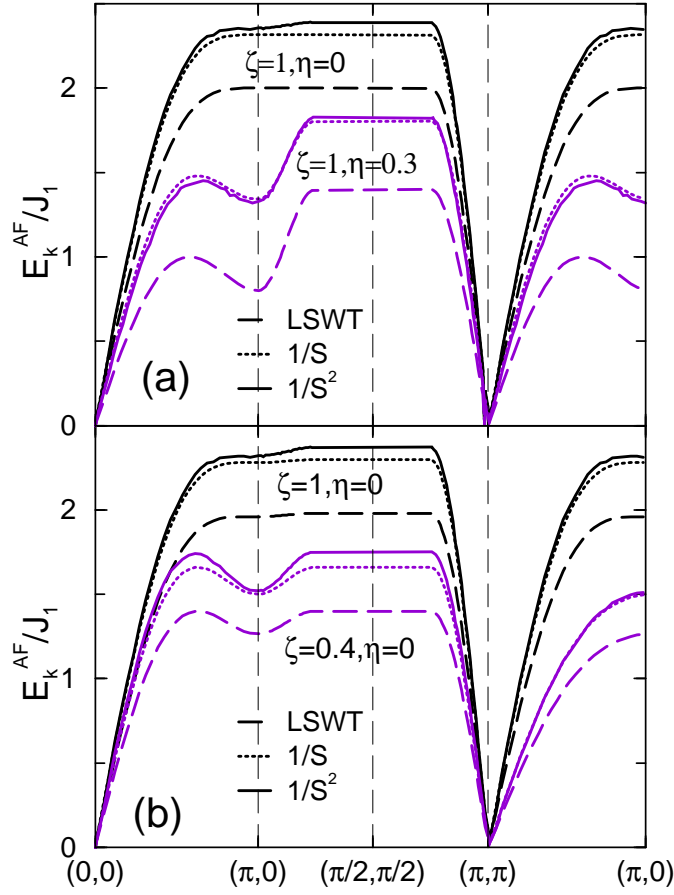}
\caption{\label{fig:CompareAFenergy} Spin-wave energy $E_{\bf k}^{\rm AF}/J_1$ obtained from  
LSWT (long-dashed lines), with first-order (dotted lines) and second-order corrections (solid lines)
for the AF-ordered phase. 
Figure (a) is for isotropic
coupling $\zeta=1$ with $\eta=0$ and $\eta=0.3$ and Fig.~(b) is 
for $\eta=0$ (no frustration) and with two different values of anisotropy
parameter $\zeta=1$ and $0.4$. Both the first and second
order corrections make the spin-wave energy larger and the corrections 
from LSWT (long-dashed lines) are 
significant in the entire BZ. (color online)}
\end{figure}

Figure~\ref{fig:CompareAFenergy} shows a comparison between the results from LSWT (long-dashed lines),
first-order (dotted lines) and second-order corrections (solid lines) to the spin-wave energy. 
Fig.~\ref{fig:CompareAFenergy}(a) shows the spin-wave energies for isotropic
coupling ($\zeta=1$) with $\eta=0$ and $\eta=0.3$ and 
Fig.~\ref{fig:CompareAFenergy}(b) 
shows the results with two different values of anisotropy
parameter $\zeta=1$ and $0.4$ for $\eta=0$. We find that in the entire BZ both the first (dotted lines) and second order corrections (solid lines) make the spin-wave energy larger and the corrections 
from LSWT (long-dashed lines) are significant for all cases. 

\begin{figure}[httb]
\centering
\includegraphics[width=3.5in]{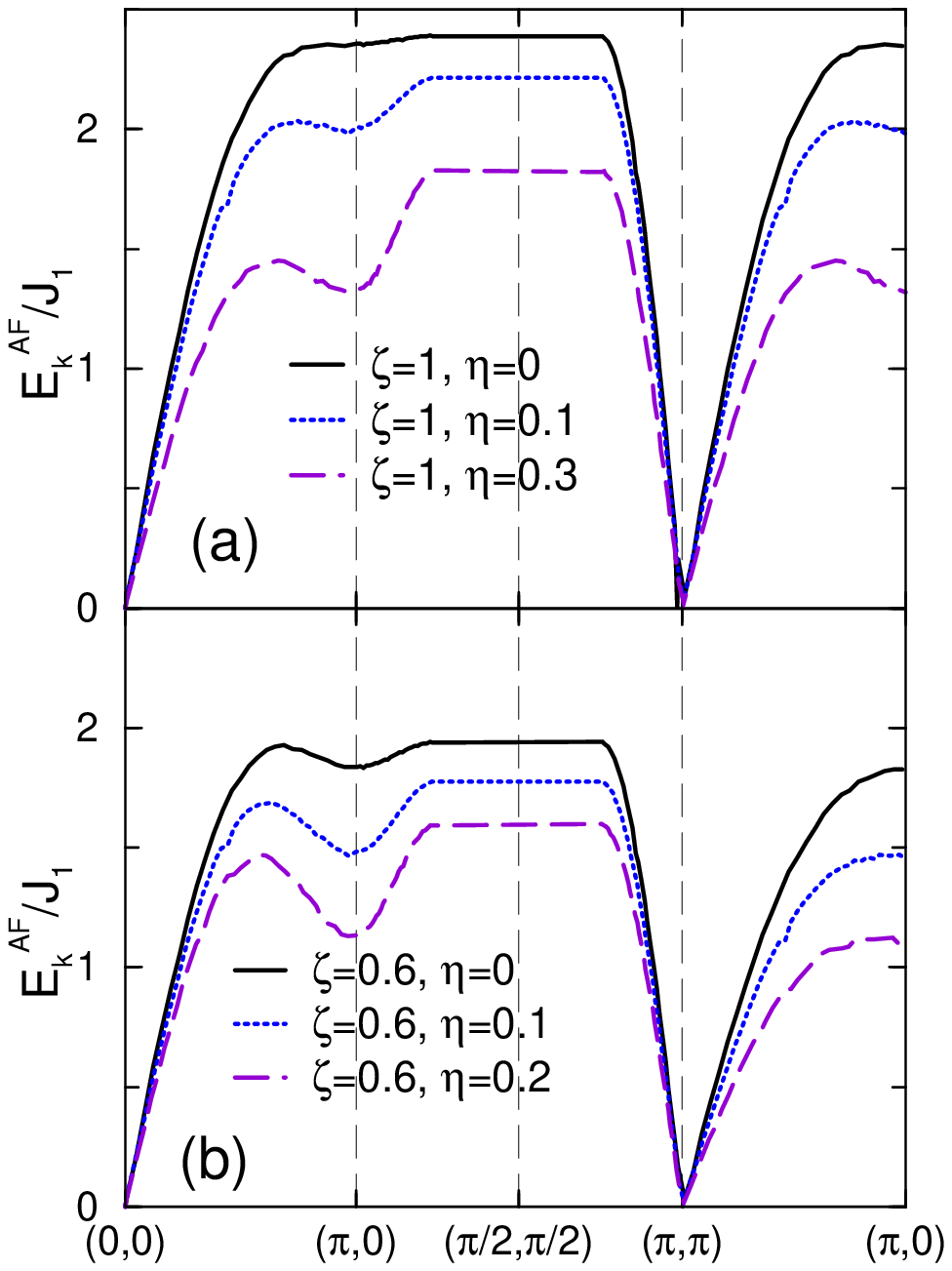}
\caption{\label{fig:AFenergydisp} Spinwave energy $E_{\bf k}^{\rm AF}/J_1$ for the AF ordered 
phase with second-order corrections is plotted for different values of $\zeta$ and $\eta$. 
The dispersion along 
($\pi/2,\pi/2$)--($\pi,0$) is flat within LSWT and 1/S correction.
The second-order corrections
make the excitation energies at $(\pi,0)$ smaller than the energies at $(\pi/2,\pi/2)$ for 
all cases. With increase in NNN frustration $\eta$ (for a fixed value of the directional parameter
$\zeta$) the dip in the magnon energy at ($\pi,0$) increases. This can
provide a measure of the strength of NNN frustration. (color online)}
\end{figure}

In Fig.~\ref{fig:AFenergydisp} we show the spin-wave energy results with second-order corrections
for different values of 
$\zeta$ and $\eta$. The spin-wave energy curve 
for the isotropic coupling $\zeta=1$ with $\eta=0$
was reported earlier.~\cite{igar05} The dispersion along 
($\pi/2,\pi/2$)--($\pi,0$) is flat within LSWT and 1/S correction (See Fig.~\ref{fig:CompareAFenergy}
for example). The second-order corrections
make the excitation energies at $(\pi,0)$ smaller than the energies at $(\pi/2,\pi/2)$. 
Our results for spin-wave energy with frustration and with anisotropic couplings are new. 
The dip in the magnon energy at ($\pi,0$) increases
with increase in frustration $\eta$. Experimentally this can provide a measure of the 
strength of NNN frustration.

\begin{figure}[httb]
\centering
\includegraphics[width=4.0in]{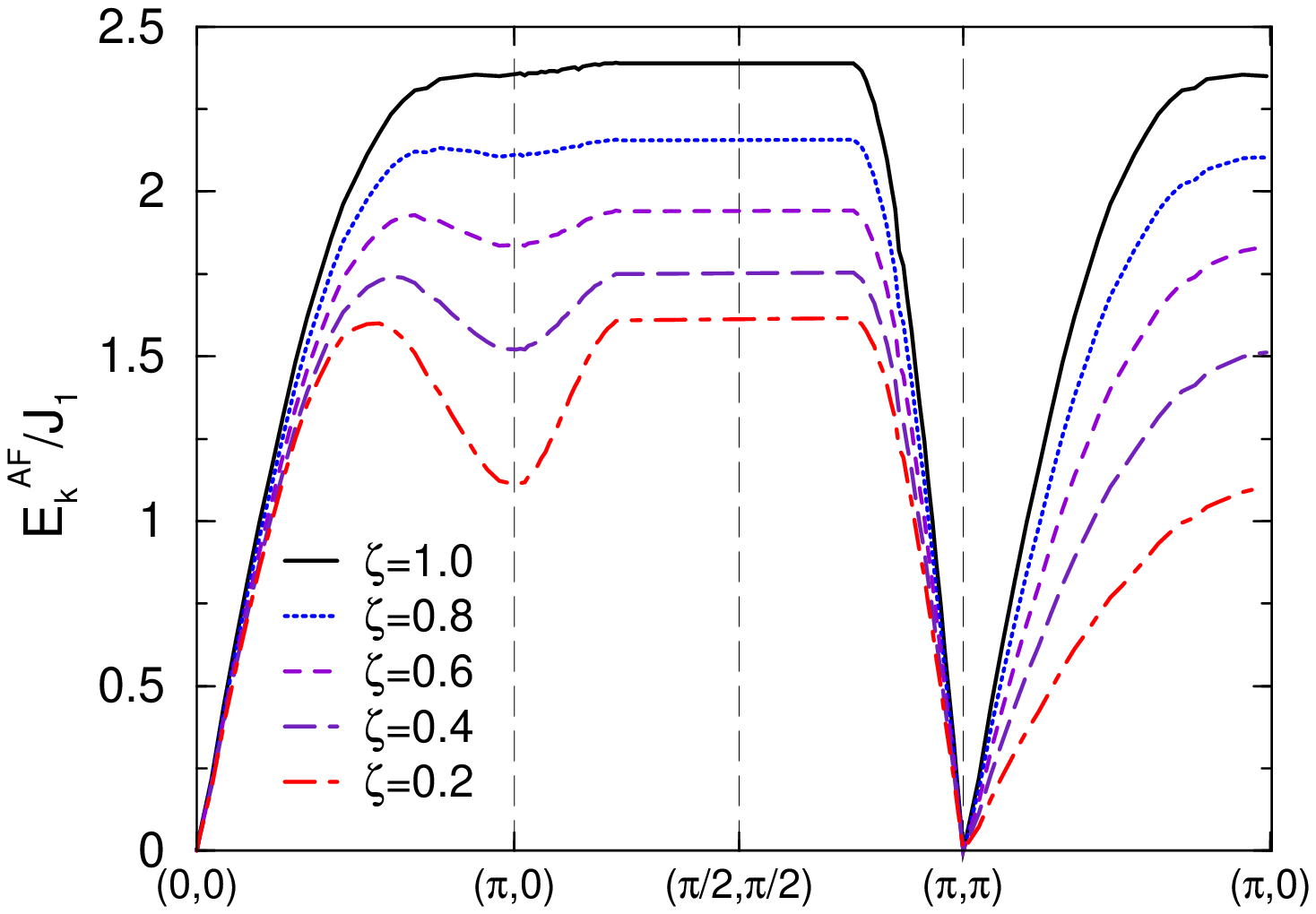}
\caption{\label{fig:AFenergy} Effect of directional anisotropy parameter 
$\zeta$ on the spin-wave energy. The calculated spin-wave energy is with second-order corrections. 
Similar to Fig.~\ref{fig:AFenergydisp} the dip
in the energy at ($\pi,0$) increases with increase in the values of $\zeta$. (color online)}
\end{figure}

In Fig.~\ref{fig:AFenergy} we show the effect of the directional anisotropy parameter 
$\zeta$ on the spin-wave energy (with second-order corrections). Similar to Fig.~\ref{fig:AFenergydisp} we find that the dip in the energy at ($\pi,0$) increasing values of $\zeta$. 

\begin{figure}[httb]
\centering
\includegraphics[width=3.5in]{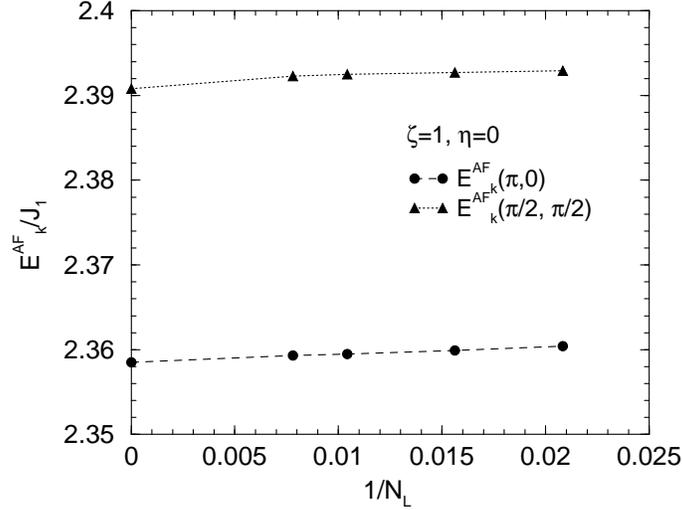}
\caption{\label{fig:Ekconv} Convergence of $E^{\rm AF}_{(\pi,0)}/J_1$ and $E^{\rm AF}_{(\pi/2,\pi/2)}/J_1$ is shown with $1/N_L$ for $N_L=48, 64,96$ and 128. We extrapolate these results using the fitting
function $A+B/N_L+C/N_L^2+D/N_L^3$ to obtain  
$E^{\rm AF}_{(\pi,0)}/J_1 = 2.3585,\; E^{\rm AF}_{(\pi/2,\pi/2)}/J_1 = 2.3908$ for $N_L \rightarrow \infty$.}
\end{figure}

Recently using neutron scattering measurements on copper deuteroformate tetradeurate (CFTD), a real two dimensional Heisenberg AF with weak interplane interactions 
($\approx 10^{-5}-10^{-4}J_1$) magnon energies have been obtained for the entire
BZ.~\cite{chris04,chris07}. It was found that the energies at ($\pi,0$) is 13.5180 meV (with estimated error of 0.1641 meV), which is about 7(1)\% smaller than the energy 14.4880 meV 
(with estimated error of 0.0647 meV) at ($\pi/2,\pi/2$).~\cite{chris07,chris} The coupling $J_1$ is estimated to be 6.19 meV. This local minimum 
at ($\pi,0$) is due to quantum fluctuations and may be due to multimagnon processes 
(entanglement of spins on neighboring sites) at this
zone boundary. 
Series expansion around the Ising limit~\cite{weihong05} and Quantum Monte 
Carlo methods~\cite{sandvik01} have accounted for all 
of the experimental data. But these numerical methods do not provide any insight into the 
physics at this zone boundary. To test our numerical procedure 
we systematically calculate the values of $E^{\rm AF}_{(\pi,0)}/J_1$ and
$E^{\rm AF}_{(\pi/2,\pi/2)}/J_1$ for $N_L=36, 48, 64,96$ and 128. The convergence of our results are very good as shown in Fig~\ref{fig:Ekconv}. We extrapolate these results using the fitting
function $A+B/N_L+C/N_L^2+D/N_L^3$ to obtain $A$ for $N_L \rightarrow \infty$ and reproduce the numerical
results $E^{\rm AF}_{(\pi,0)}/J_1 \approx 2.3585,\; E^{\rm AF}_{(\pi/2,\pi/2)}/J_1 \approx 2.3908$ 
reported earlier~\cite{igar05} with a 1.4\% decrease between these two energy values. 
Recently a third-order in 1/S expansion has been done to obtain the spectrum of short-wavelength magnons~\cite{syrom,syrom10} where it was shown that the 1/S series
converges slowly near the wave-vector $(\pi,0)$. With the third order correction the excitation
energy at ($\pi,0$) was found to be 3.2\% smaller than at ($\pi/2,\pi/2$). This result for the energy difference still falls short of the experimental result of 7\%. This suggests that the inclusion of correction to even third order in 1/S is insufficient to explain this energy difference. It should be noted that other interactions e.g. ring exchange interactions have been proposed to play a role in these compounds~\cite{coldea01,roger05,schmidt05}.

It may be interesting to study the effects of small NNN frustration and small anisotropy on the energies at these two zone boundaries. Table~\ref{table:energy2} shows our extrapolated values  of $E^{\rm AF}_{(\pi,0)}/J_1$ and $E^{\rm AF}_{(\pi/2,\pi/2)}/J_1$ and the percentage changes for small frustrations $\eta=0.01,0.02$ and a small directional anisotropy
$\zeta=0.98$. Calculations are done with lattice sizes $N_L=48,64,96$ and $128$ and the results are
extrapolated to $N_L \rightarrow \infty$ using the fitting function $A+B/N_L+C/N_L^2+D/N_L^3$. We show that a small frustration (for example $\eta=0.02$) for the isotropic coupling causes a noticeable difference (2.8\% within second-order spin-wave expansion) in energies between $E^{\rm AF}_{(\pi,0)}/J_1$ and $E^{\rm AF}_{(\pi/2,\pi/2)}/J_1$. These features can be explored experimentally using neutron scattering measurements with compounds that can be modeled by the $J_1-J_1^\prime-J_2$ Heisenberg antiferromagnet.  
\begin{table}[htbp]
 \caption{Energies $E^{\rm AF}_{(\pi,0)}/J_1$ and $E^{\rm AF}_{(\pi/2,\pi/2)}/J_1$ for different values
of $\zeta$ and $\eta$.}
\centering
\newcommand{\mc}[3]{\multicolumn{#1}{#2}{#3}}
\begin{tabular}[c]{||l|l|l|l|l||}\hline
 $N_L \rightarrow \infty$ & × & $E^{\rm AF}_{(\pi,0)}/J_1$  & $E^{\rm AF}_{(\pi/2,\pi/2)}/J_1$ & 
$\Delta E^{\rm AF}/E^{\rm AF}_{(\pi/2,\pi/2)}$ \\ 
\hline \hline
\mc{1}{|c|}{×} & \mc{1}{c|}{LSWT} & \mc{1}{c|}{2.0000} & \mc{1}{c|}{2.0000} & \mc{1}{c|}{0\%} \\ 
\hline
\mc{1}{|c|}{$\zeta=1,\eta=0$} & \mc{1}{c|}{1/S} & \mc{1}{c|}{2.3159} & \mc{1}{c|}{2.3159} & \mc{1}{c|}{0\%} \\ 
\hline
\mc{1}{|c|}{×} & \mc{1}{c|}{1/S$^2$} & \mc{1}{c|}{2.3585} & \mc{1}{c|}{2.3908} & \mc{1}{c|}{1.4\%} \\ 
\hline \hline
\mc{1}{|c|}{×} & \mc{1}{c|}{LSWT} & \mc{1}{c|}{1.9600} & \mc{1}{c|}{1.9800} & \mc{1}{c|}{1.0\%} \\ 
\hline
\mc{1}{|c|}{$\zeta=1,\eta=0.01$} & \mc{1}{c|}{1/S} & \mc{1}{c|}{2.2800} & \mc{1}{c|}{2.2980} & \mc{1}{c|}{0.8\%} \\ 
\hline
\mc{1}{|c|}{×} & \mc{1}{c|}{1/S$^2$} & \mc{1}{c|}{2.3221} & \mc{1}{c|}{2.3753} & \mc{1}{c|}{2.2\%} \\ 
\hline \hline
\mc{1}{|c|}{×} & \mc{1}{c|}{LSWT} & \mc{1}{c|}{1.9200} & \mc{1}{c|}{1.9600} & \mc{1}{c|}{2.0\%} \\ 
\hline
\mc{1}{|c|}{$\zeta=1,\eta=0.02$} & \mc{1}{c|}{1/S} & \mc{1}{c|}{2.2443} & \mc{1}{c|}{2.2801} & \mc{1}{c|}{1.6\%} \\ 
\hline
\mc{1}{|c|}{×} & \mc{1}{c|}{1/S$^2$} & \mc{1}{c|}{2.2886} & \mc{1}{c|}{2.3536} & \mc{1}{c|}{2.8\%} \\ 
\hline \hline
\mc{1}{|c|}{×} & \mc{1}{c|}{LSWT} & \mc{1}{c|}{1.9799} & \mc{1}{c|}{1.9800} & \mc{1}{c|}{0\%} \\ 
\hline
\mc{1}{|c|}{$\zeta=0.98,\eta=0$} & \mc{1}{c|}{1/S} & \mc{1}{c|}{2.2926} & \mc{1}{c|}{2.2928} & \mc{1}{c|}{0\%} \\ 
\hline
\mc{1}{|c|}{×} & \mc{1}{c|}{1/S$^2$} & \mc{1}{c|}{2.3348} & \mc{1}{c|}{2.3680} & \mc{1}{c|}{1.4\%} \\ 
\hline \hline
\end{tabular}
\label{table:energy2}
\end{table}

\subsubsection{\label{sec: AFphase-velocity}Renormalized spin-wave velocities}
\begin{figure}[httb]
\centering
\includegraphics[width=3.5in]{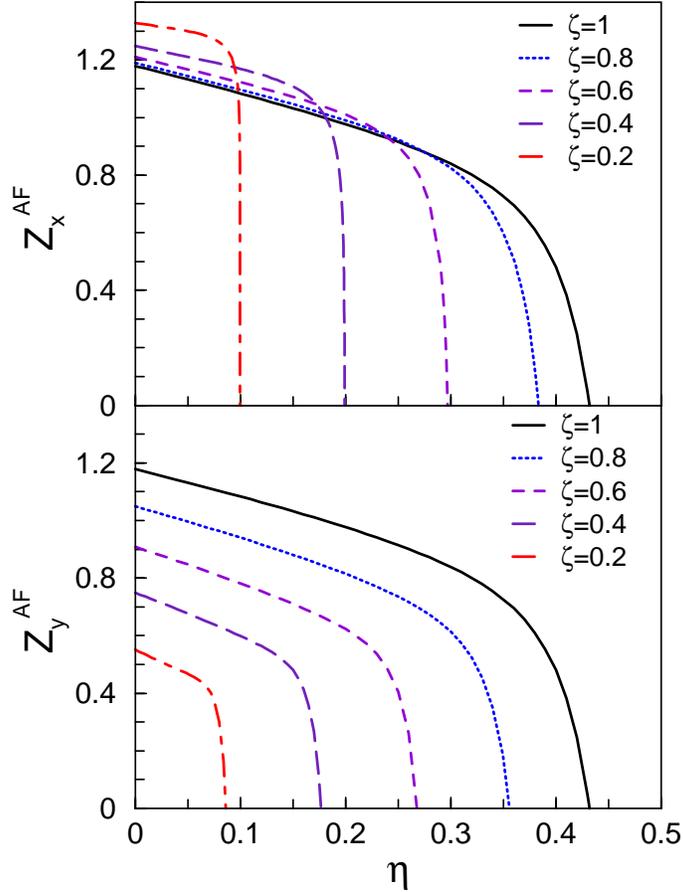}
\caption{\label{fig:vel-AF} Renormalization factors $Z_{v_x}^{\rm AF}$ and $Z_{v_y}^{\rm AF}$ for the spin-wave velocities are
plotted with frustration $\eta$ for different values of $\zeta$. The velocities steadily decrease
with increase in NNN frustration and finally becomes zero close to the quantum critical 
transition points $\eta_{1c}$ for the AF-phase.
(color online)}
\end{figure}

We calculate the spin-wave velocity renormalization factors $Z_{v_x}^{\rm AF}, Z_{v_y}^{\rm AF}$ along the $x$ and $y$ directions from Eq.~\ref{vel-AF}. For the second-order correction terms $v_{2x}, v_{2y}$, we consider lattice 
size $N_L=72$ and  evaluate 
$[\Sigma^{(2)}_{\alpha \alpha}({\bf k},E_{\bf k})-
\frac {B_{\bf k}^2}{2E_{\bf k}}]/k_x$ with $k_x=\pi/N_L$. $v_{2y}$ is obtained similarly. For the isotropic
case $\zeta=1$ and with $\eta=0$ we find the second-order correction $v_{2x}=v_{2y}=0.021$ which is in 
excellent agreement with results reported earlier.~\cite{canali92A,igar93,igar05}
The results from our calculations with increase in $\eta$ are shown in Fig.~\ref{fig:vel-AF}. We find that the velocities steadily decrease with increase in frustration and finally becomes zero close to the quantum critical points $\eta_{1c}$ for the AF-phase. Second-order corrections are significant to stabilize the velocities as with first-order corrections these velocities diverge with increase in frustration (similar to the case with magnetization discussed later). We also notice that the difference between 
the renormalization factors $Z_{v_x}^{\rm AF}$ and $Z_{v_y}^{\rm AF}$ diminishes with increase 
in frustration. 

\subsubsection{\label{sec: AFphase-mag}Staggered Magnetization}
\begin{figure}[httb]
\centering
\includegraphics[width=4.5in]{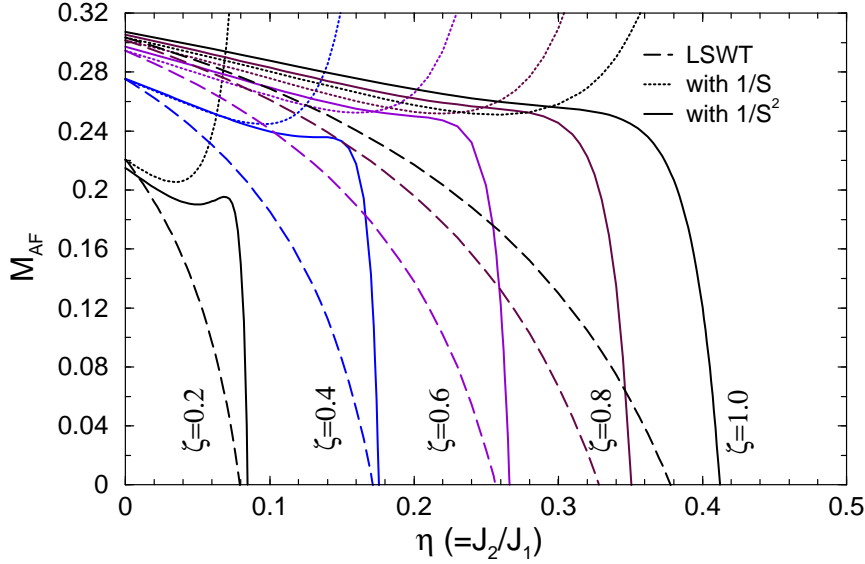}
\caption{\label{fig:AFmag} Staggered magnetization $M_{\rm AF}$ 
is shown for the AF ordered phase with frustration $\eta = J_2/J_1$ with different values
of spatial anisotropy $\zeta = J_1^\prime/J_1$. For each
$\zeta$ results from LSWT (long-dashed lines),
with first-order (dotted lines) and with second-order corrections (solid lines) 
are plotted. With increase in $\eta$
the dotted curves diverge. Second-order 1/S$^2$ corrections become significant 
for large $\eta$ and they stabilize the apparent divergence of the magnetization.
Magnetizations with 1/S$^2$ corrections decrease steadily and then sharply drop to zero. 
For example in the isotropic case i.e. $\zeta=1$, 
$M_{\rm AF}$ starts from 0.307 and then decreases till $\eta \approx 0.32$ 
and finally becomes zero
at the critical point $\eta_{1c} \approx 0.41$. However for small $\zeta$, say 
$\zeta=0.2$ we find $M_{\rm AF}$ to steadily decrease 
from 0.21 at $\eta=0$ to 0.19 at $\eta \approx 0.054$ and then slightly increases 
to 0.195 at $\eta \approx 0.068$. Finally it
sharply drops to zero at $\eta_{1c} \approx 0.084$. Note that for all cases second-order 
corrections increase the critical value of $\eta_{1c}$ from the LSWT predictions. 
(color online)}
\end{figure}

We obtain the staggered
magnetization $M_{\rm AF}$ for the AF phase with several values of $\zeta$ and
$\eta$ from Eq.~\ref{Mag-AF} by numerically evaluating Eqs.~\ref{MagAF-LSWT}--\ref{M2-AF}. 
Especially to obtain the second order correction term $M_2$ we sum
up the values of $N_L^2/4$ points of ${\bf k}$ in the 1/4-part of the 
first BZ and $N_L^2$ points of ${\bf p}$ and ${\bf q}$ in the first BZ,
with $N_L=36$ lattice sites (total of about 544.2 million points for each $\zeta$ and $\eta$). 
Except for small spatial anisotropy $\zeta$, $M_2$ values 
start from a positive small 
number and then switch sign and become negative with increase in 
frustration $\eta$. However,
for small $\zeta$, say $\zeta=0.2$ $M_2$ starts from a small negative number 
($\sim -0.005$) and remains
negative with increase in $\eta$. Figure~\ref{fig:AFmag} shows the magnetization with 
increase in the frustration parameter $\eta=J_2/J_1$ for several values of the spatial
anisotropy parameter $\zeta = J_1^\prime/J_1 = 0.2, 0.4,0.6,0.8$, and $1.0$. For each
$\zeta$ three different curves are plotted: the long-dashed lines represent LSWT prediction,
the dotted lines include the first-order ($1/S$) correction to the LSWT results, and the solid lines 
represent corrections up to second-order ($1/S^2$) to the LSWT results. With increase in frustration
the dotted curves diverge. However, 1/S$^2$ corrections ($M_2$) significantly increase
with frustration and stabilize the apparent divergence of the magnetization.
We find that the magnetization with second-order corrections decreases steadily at first 
and then sharply drops to zero. As an example, for the isotropic case ($\zeta=1, \eta=0$), 
$M_{\rm AF}$ starts
from 0.307 and then decreases till $\eta \approx 0.32$ 
and finally becomes zero
at the critical point $\eta_{1c} \approx 0.41$. For this case we reproduce the 
magnetization plot obtained in Ref.~\onlinecite{igar92}. Other values of $\zeta$ show the same
trend except for small $\zeta$. For $\zeta=0.2$ we find $M_{\rm AF}$ steadily decreases from 
0.21 at $\eta=0$ to 0.19 at $\eta \approx 0.054$ and then slightly increases to 0.195 at $\eta \approx 0.068$.
Finally it sharply drops to zero at $\eta_{1c} \approx 0.084$. This feature has not been observed before and may be
an artifact of the spin-wave expansion showing the limitation of this method for small 
$\zeta$ (the system becomes essentially one-dimensional as $\zeta \rightarrow 0$). It may be 
interesting to verify this by series expansion or other analytical or numerical methods. Note that for
all cases second-order corrections increase the critical value of $\eta_{1c}$ from the
LSWT predictions.
Our values of magnetization for $\eta=0$ (no NNN frustration), $M_{\rm AF}^{(0)}=0.307$ 
agrees with previously obtained
values from spin-wave expansion~\cite{huse88,igar92,igar93,igar05}, 
series expansion~\cite{weihong91-2,hamer92,weihong91,weihong93,oitmaa96}, and 
experimental results for K$_2$NiF$_4$,
K$_2$MnF$_4$, Rb$_2$MnF$_4$ and other systems~\cite{kim99,coldea01,wijn70a,wijn70b,chris07}.

\begin{figure}[httb]
\centering
\includegraphics[width=4.0in]{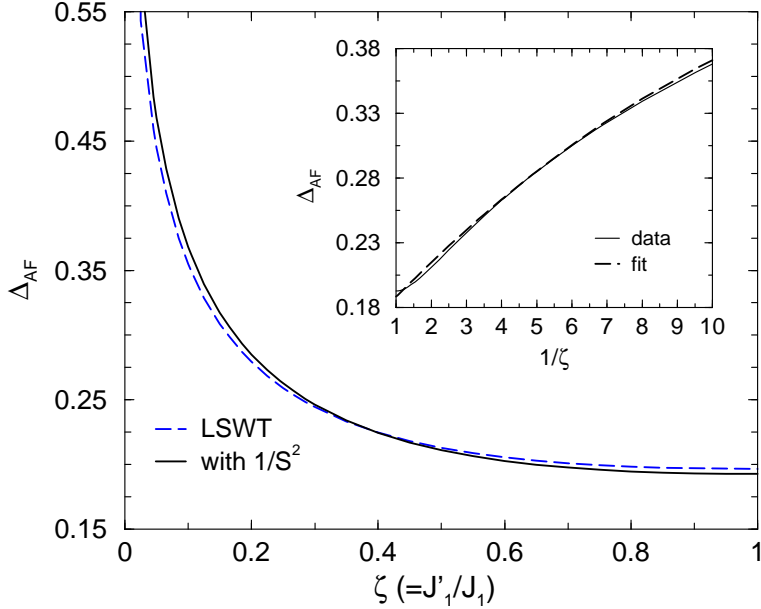}
\caption{\label{fig:M0AF} Spin deviation  $\Delta_{\rm AF}=0.5- M_{\rm AF}^{(0)}$ from the classical value of 0.5
is plotted for the AF ordered phase (with no NNN interaction, i.e. $\eta=0$) for 
different values of spatial anisotropy $\zeta$. Dashed line is LSWT prediction
whereas  the solid line includes $1/S^2$ corrections to LSWT results. The fluctuation 
increases with decreasing values of $\zeta$, suggesting that spin-wave expansion for 
$S=1/2$ is unreliable for $\zeta < 0.1$. In the inset we show the fluctuations 
with $1/\zeta$ for $\zeta=0.1$ to 1. The function
$f(\zeta) = 0.16+0.029\zeta^{-1}-0.00079\zeta^{-2}$ 
is a good representation of $\Delta_{\rm AF}$ for $\zeta=0.1 - 1.0$. The fitted curve
is shown in the inset along with the actual numerical results with the second-order 
corrections. (color online)}
\end{figure}

The ground-state magnetization per spin is reduced from its classical value
$S=1/2$  by zero-point quantum fluctuations. This ``spin reduction'' 
$\Delta_{\rm AF}=0.5- M_{\rm AF}^{(0)}$ is plotted for different values of $\zeta$ 
for $\eta=0$ in
Fig.~\ref{fig:M0AF}. Second-order corrections (solid line) change the values of 
$M_{\rm AF}^{(0)}$ slightly from the LSWT predictions (dashed line). The fluctuations 
increase with decreasing values of $\zeta$, suggesting that spin-wave expansion for 
$S=1/2$ is not applicable for $\zeta < 0.1$ as the system essentially becomes one 
dimensional.

In the inset of Fig.~\ref{fig:M0AF} we show the spin deviation with $1/\zeta$ for $\zeta=0.1$ to
1. We find that the function
$f(\zeta) = 0.16+0.029\zeta^{-1}-0.00079\zeta^{-2}$ 
is a good representation of $\Delta_{\rm AF}$ for this range of $\zeta$. The fitted curve (dashed line)
is shown in the inset along with the actual numerical results (solid line) 
with the second-order corrections.

\subsection{\label{sec: CAFphase-result}CAF Phase}
\subsubsection{\label{sec: CAFphase-energy}Spin-wave energy dispersion}

\begin{figure}[httb]
\centering
\includegraphics[width=3.5in]{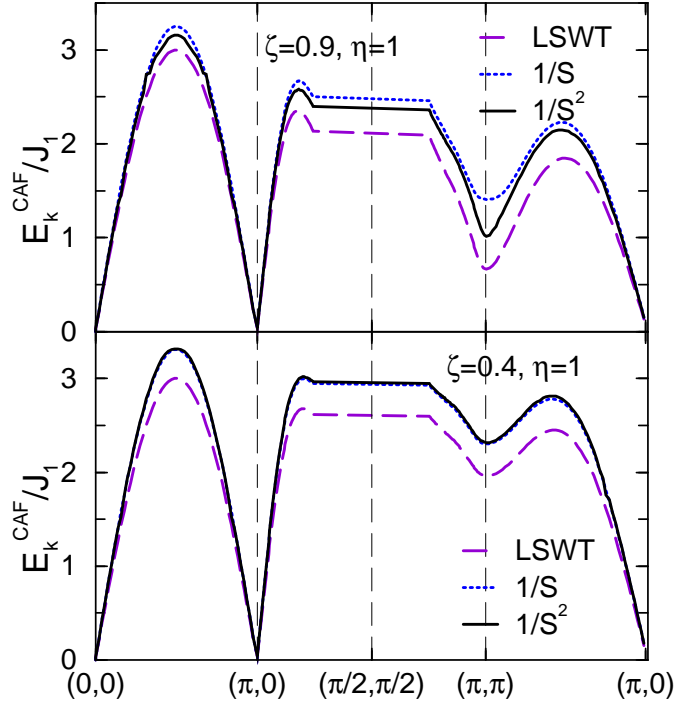}
\caption{\label{fig:CompareCAFenergy} Spin-wave energy $E^{\rm CAF}_{\bf k}/J_1$ 
results obtained from linear spin-wave theory 
(long-dashed lines), with first-order (dotted lines), and second-order corrections (solid lines)
for the entire Brillouin zone of the CAF-ordered phase. For fixed NNN frustration $\eta=1$ two different values of $\zeta=0.9$ and 0.4 are chosen. 1/S and 1/S$^2$ corrections increase the spin-wave energy
of the ordered phase from the linear-spin wave theory results. For $\zeta=0.4$ second order corrections are insignificant compared to the first order 1/S corrections. However, for $\zeta=0.9$ 1/S$^2$ corrections lower the spin-wave energy from the first-order
corrections. Spin-wave energy shows three peaks, the maximum being at $(\pi/2,0)$. The second small peak is at $(0.514\pi, 0.486\pi)$ and the third peak occurs at $(\pi,\pi/2)$. (color online)}
\end{figure}
We numerically evaluate Eq.~\ref{energyEk-CAF} with $N_L=72$ lattice size to obtain the spin-wave energy 
$2J_1S(1+2\eta){\tilde E_{\bf k}^{\rm CAF}}$ as a function of momentum for several values
of $\zeta$ and $\eta$. The calculations are similar to the AF-phase. Figure~\ref{fig:CompareCAFenergy} shows the spin-wave energy ${\tilde E^{\rm CAF}_{\bf k}}/J_1$ 
with second-order corrections (solid lines) for the entire Brillouin zone of the CAF-ordered phase.
Two different values of directional anisotropy $\zeta=0.9$ and 0.4 for NNN frustration $\eta=1$ are chosen. Results 
obtained from linear spin-wave theory (long-dashed lines) and with only first-order (dotted lines) are 
also shown for comparison. 1/S and 1/S$^2$ corrections increase the energy
of the ordered phase from the LSWT results. We find that the second-order corrections to the magnon 
energy are not significant from the energy obtained with first-order corrections for small $\zeta$. However for large $\zeta$, say $\zeta=0.9$ 1/S$^2$ corrections lower the spin-wave energy from the first-order 1/S corrections.

\begin{figure}[httb]
\centering
\includegraphics[width=4.0in]{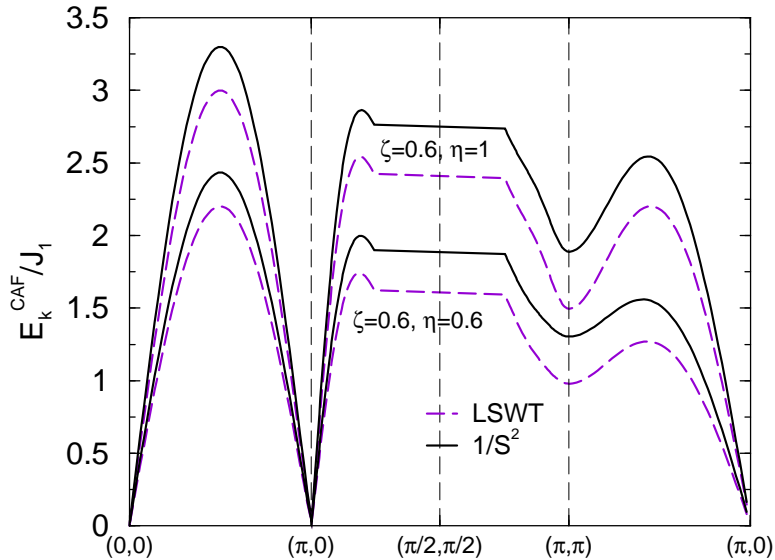}
\caption{\label{fig:CAFenergy} Effect of frustration 
$\eta$ on the spin-wave energy in the CAF-phase. Second-order corrections are negligible compared to the first-order corrections. However, 1/S corrections significantly enhance the spin-wave energy obtained from LSWT results. (color online)}
\end{figure}

Figure~\ref{fig:CAFenergy} shows the effect of frustration $\eta$ for a fixed value of spatial 
anisotropy $\zeta=0.6$. Second-order corrections are negligible compared to the first-order corrections,
which significantly enhance the LSWT results. In both Fig.~\ref{fig:CompareCAFenergy} and Fig.~\ref{fig:CAFenergy} the spin-wave energy vanishes at the wave-vector ($\pi,0$) as expected for the
CAF phase. We find three peaks in the magnon energy, the maximum being at $(\pi/2,0)$. The second small peak in energy is at $(0.514\pi, 0.486\pi)$ and the third peak occurs at $(\pi,\pi/2)$. 
\subsubsection{\label{sec: CAFphase-velocity}Renormalized velocities}
\begin{figure}[httb]
\centering
\includegraphics[width=3.50in]{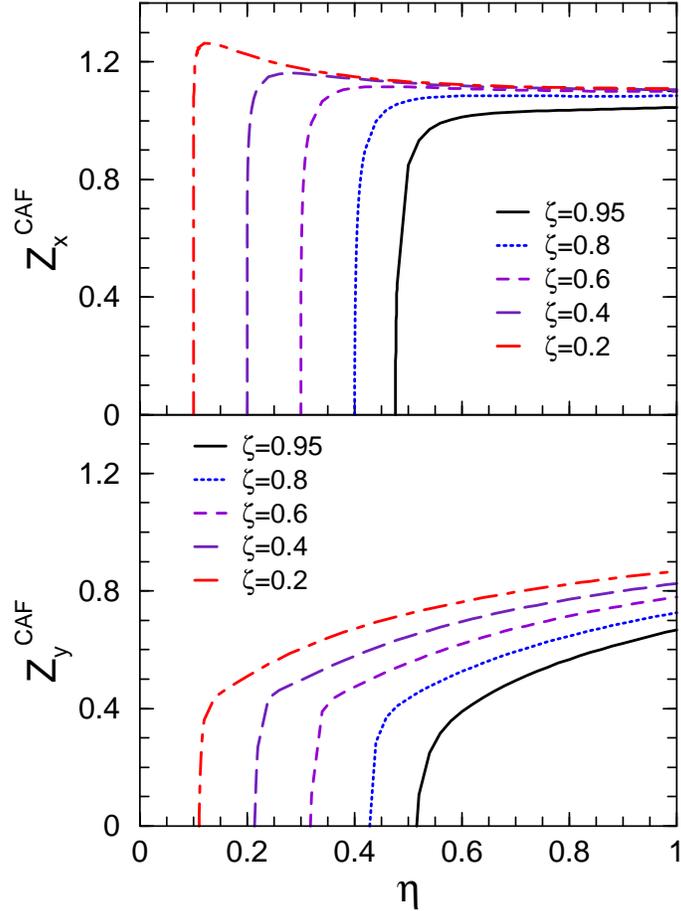}
\caption{\label{fig:vel-CAF} Renormalization factors $Z_{v_x}^{\rm CAF} $ and $Z_{v_y}^{\rm CAF}$ are
plotted with frustration $\eta$ for different values of $\zeta$. The velocities steadily decrease
with increase in NNN frustration and finally become vanish close to the quantum critical 
transition points $\eta_{2c}$ for the CAF-phase. Numerical calculations are done with lattice
size $N_L=72$. For $\zeta=0.2,0.4$ we find that $Z_{v_x}^{\rm CAF}$ slightly increases and then sharply drops to zero. Increasing the lattice size to $N_L=96$ does not change our results. (color online)}
\end{figure}

Renormalization factors $Z_{v_x}^{\rm CAF}, Z_{v_y}^{\rm CAF}$ along the $x$ and $y$ directions are obtained from Eq.~\ref{vel-CAF} with  second-order corrections. The results are shown in Fig.~\ref{fig:vel-CAF}. As we expect similar to the AF-phase
the velocities steadily decrease 
with increase in $\eta$ and finally becomes zero close to the quantum transition points $\eta_{2c}$ for the CAF-phase. Second-order corrections are significant to stabilize the velocities as with first-order corrections these velocities diverge with increase in frustration (similar to the case with magnetization discussed later). For small $\zeta$ ($\zeta=0.2,0.4$ in figure) we find that $Z_{v_x}^{\rm CAF}$ slightly increases and then sharply drops to zero. We increase the lattice size to $N_L=96$ to check the accuracy
of our calculation. We find no changes in our plot. It may be interesting to verify this with series
expansion or other analytical or numerical methods. Our numerical method based on the spin-wave 
expansion for the CAF phase is not reliable for $\zeta>0.95$ -- so we have not been able to obtain the 
renormalized spin-wave velocities for the case with $\zeta=1$ and $\eta=1$ (more discussed in Sec.~\ref{sec: CAFphase-mag}). 

\subsubsection{\label{sec: CAFphase-mag}Staggered Magnetization}
\begin{figure}[httb]
\centering
\includegraphics[width=4.5in]{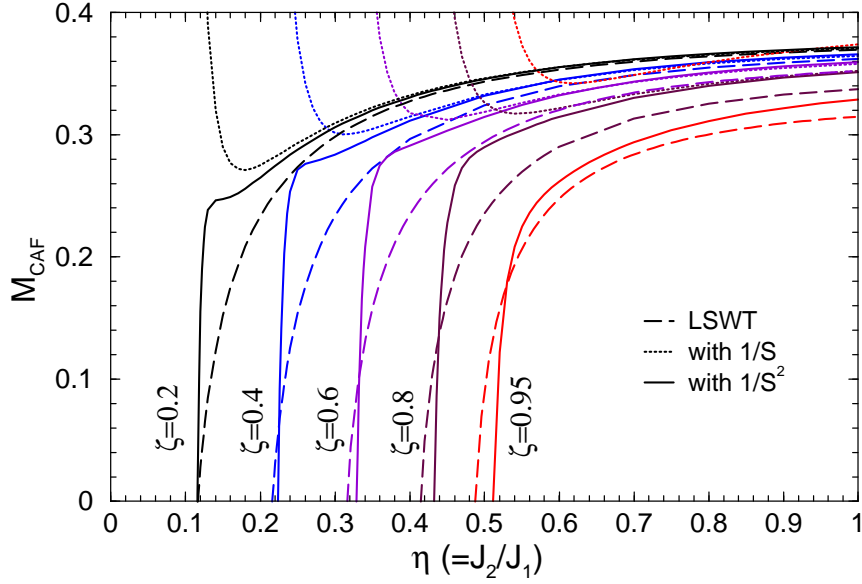}
\caption{\label{fig:CAFmag} Staggered magnetization $M_{\rm CAF}$ for the CAF ordered 
phase is plotted with $\eta$ for different values of $\zeta$. For each value of 
$\zeta$ three different curves are shown: long-dashed line is the prediction from 
LSWT, dotted line is the first-order correction, and the solid line 
includes corrections up to second-order. In all cases first-order corrections 
diverge for some
value of $\eta$. However, second-order (1/S$^2$) corrections become significant and 
stabilize the magnetization. Similar to the AF-phase $M_{\rm CAF}$ with second order 
corrections 
decreases steadily and then sharply drops to zero. 
For example with $\zeta=0.2$, 
$M_{\rm CAF}$ starts from 0.371 at $\eta=1$ and then decreases till $\eta \approx 0.12$ 
and sharply drops to zero
at the critical point $\eta_{2c} \approx 0.116$. For $\zeta$ more than 0.95 
the fluctuations become too large (see Fig.~\ref{fig:M0CAF}) -- in that case our spin-wave
expansion becomes invalid (see text). With increase in $\zeta$ the values of the critical 
points $\eta_{2c}$ 
differ more from the LSWT predictions.(color online)}
\end{figure}

Similar to the AF-phase the staggered
magnetization $M_{\rm CAF}$ for the CAF phase with several values of $\zeta$ and
$\eta$ are obtained by summing over points in the 
first BZ with $N_L=36$ lattice sites. 
Except for large spatial anisotropy $\zeta$, $M_2$ values 
start from a small positive 
number and then switches sign and become negative with increase in  
frustration $\eta$. However,
for large $\zeta$, say for $\zeta=0.8$ $M_2$ corrections are always negative. 
Figure~\ref{fig:CAFmag} shows the magnetization with 
increase in frustration parameter $\eta$ for several values of 
$\zeta=0.2, 0.4,0.6,0.8,0.95$. For each
$\zeta$ three different curves are plotted: LSWT results (long-dashed line),
first-order corrections (dotted line), and second-order corrections (solid line) 
to the LSWT results. Similar to the AF-phase 
the dotted curves diverge with increase in frustration. However, 
1/S$^2$ corrections ($M_2$) significantly increase
with frustration and stabilize the magnetization and finally make it zero.
We find that $M_{\rm CAF}$ decreases steadily at first 
and then sharply drops to zero. As an example, for the $\zeta=0.2$ 
$M_{\rm CAF}$ starts
from 0.371 and then decreases till $\eta \approx 0.12$ 
and sharply drops to zero
at the critical point $\eta_{2c} \approx 0.116$. 
With increase in $\zeta$ the values of the critical 
points $\eta_{2c}$ differ more from LSWT predictions.


\begin{figure}[httb]
\centering
\includegraphics[width=4.0in]{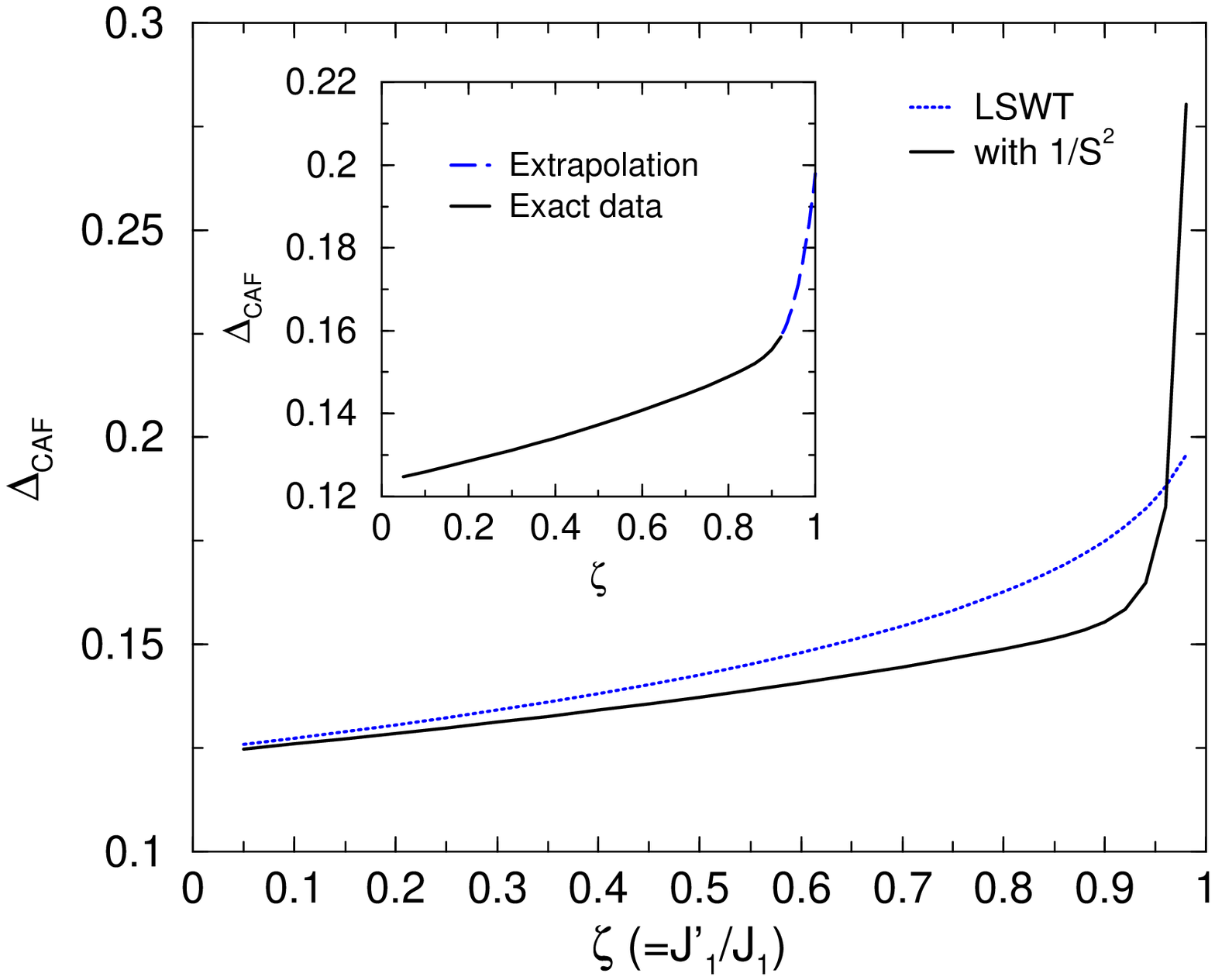}
\caption{\label{fig:M0CAF} Spin deviation $\Delta_{\rm CAF}$ 
is plotted for the CAF ordered phase ($\eta=1$). Dashed line is LSWT results
and the solid line is with the $1/S^2$ corrections. For
$\zeta$ more than 0.95 spin-wave expansion becomes unreliable as $\Delta_{\rm CAF}$ 
as the first and second-order corrections to $\Delta_{\rm CAF}$ increase rapidly. 
Thus with the 1/S$^2$ corrections we extrapolated the values to obtain 
$\Delta_{\rm CAF}\approx 0.20$ 
for $\zeta=1$. This gives $M_{\rm CAF} \approx 0.30$ for the isotropic limit. This is in 
good agreement with existing experimental results (see text).
Inset shows both the exact data (solid line) and the extrapolated curve
(dashed line). (color online)}
\end{figure}

We also find that starting from $\zeta=0.95$ the spin deviation  $\Delta_{\rm CAF}=0.5-M_{\rm CAF}^{(0)}$ 
increases substantially as we approach the isotropic limit $\zeta=1$. This is shown in 
Fig.~\ref{fig:M0CAF}. $\Delta_{\rm CAF}$ from the 
LSWT theory remains smooth (dashed lines in Fig.~\ref{fig:M0CAF}). Both  
the first ($M_1$) and second order ($M_2$) corrections 
increase rapidly for $\zeta >0.95$. This increase is 
due to the fact that $\epsilon^\prime_{\bf k} \rightarrow 0$ as $\zeta \rightarrow 1$. 
We have not found a numerical way to regulate it. Instead we used extrapolation 
to obtain values of $\Delta_{\rm CAF}$ beyond $\zeta=0.95$. Inset of 
Fig.~\ref{fig:M0CAF} shows both the exact data (solid line) and the extrapolated curve
(dashed line). With the extrapolated curve we obtain $\Delta_{\rm CAF}\approx 0.20$ 
for $\zeta=1$, which gives $M_{\rm CAF} \approx 0.30$ for the isotropic limit.
This is in good agreement with the recent neutron scattering measurements data of 
the order parameter $M_{\rm CAF}=0.31(2)$ for Li$_2$VOSiO$_4$, 
which is believed to be a $S=1/2$ frustrated antiferromagnet on a 
square lattice with $J_2 \approx J_1$.~\cite{melzi00,melzi01,carretta02,bombardi04}

\subsection{\label{sec: AF-CAF}AF and CAF ordered phases - Phase Diagram}
\begin{figure}[httb]
\centering
\includegraphics[width=4.5in]{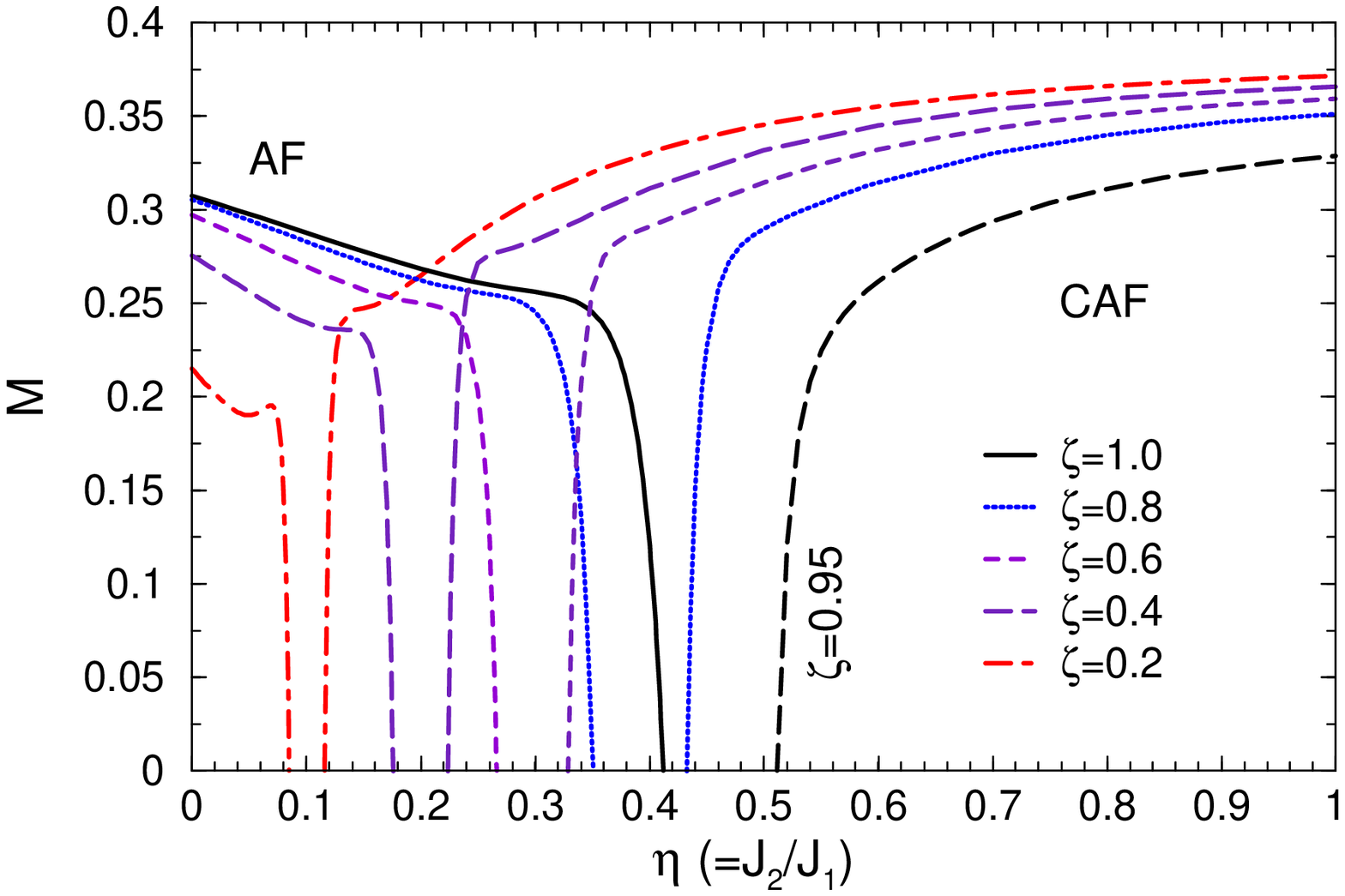}
\caption{\label{fig:AFCAFmag} Staggered magnetization $M$
is plotted for both the AF and CAF ordered phase (with second-order corrections). For the 
AF phase the different values of $\zeta$ are 0.2, 0.4, 0.6, 0.8, and 1.0 and for the 
CAF phase the values are 0.2, 0.4, 0.6, 0.8, and 0.95. Our numerical approach using spin-wave expansion 
is not 
reliable for the CAF phase for $\zeta$ larger than 0.95 (see text for details). For both
the phases $M$ become zero at some critical values of the NNN frustration parameter $\eta$.
We also find that the spin-gap increases with increase in $\eta$. (color online)}
\end{figure}

Staggered magnetizations of a spatially anisotropic frustrated 
spin-1/2 Heisenberg antiferromagnet on a square lattice is presented in 
Fig.~\ref{fig:AFCAFmag} for both AF and CAF ordered phases. We find the staggered magnetization $M^{(0)}_{\rm AF}$ 
for $\eta=0$ to decrease 
with decrease in anisotropy $\zeta$ in the AF-phase. On the other hand, $M^{(0)}_{\rm CAF}$ increases 
with decrease in $\zeta$ for $\eta=1$. Our results for $M^{(0)}_{\rm AF}=0.307$ for the AF-ordered
phase and
$M^{(0)}_{\rm CAF}=0.30$ for the CAF-ordered phase are in excellent agreement with existing 
experimental data on these systems. Furthermore, we find that in both the phases the second
order corrections play a significant role to stabilize the magnetization. Staggered magnetizations
become zero in both the phases at the critical values $\eta_{1c}$ and $\eta_{2c}$ for each
value of $\zeta$. 

\begin{figure}[httb]
\centering
\includegraphics[width=3.5in]{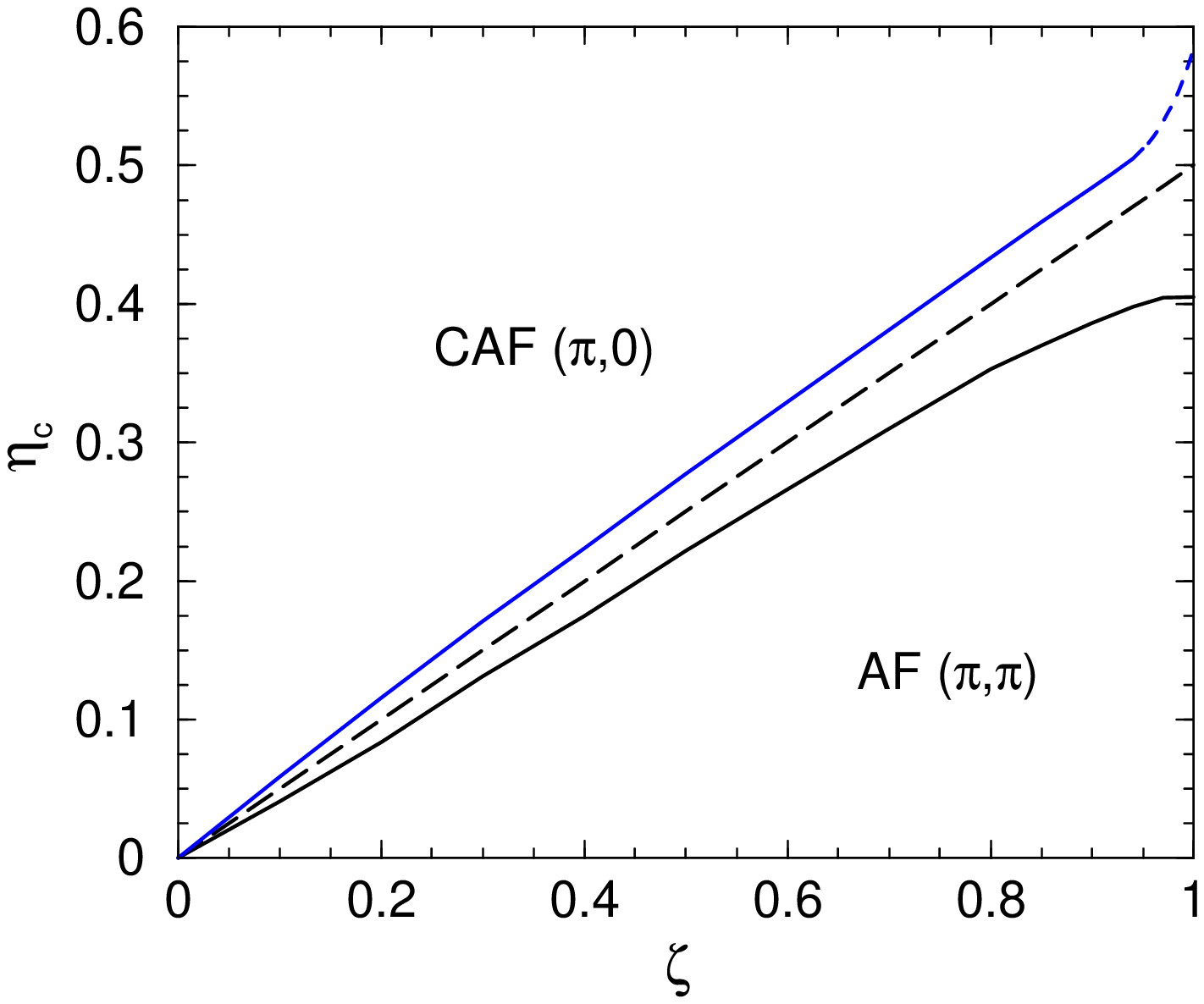}
\caption{\label{fig:critical} Phase diagram for the $J_1-J_1^\prime-J_2$ model.
The solid lines indicate the critical points $\eta_{1c}$ for the AF and 
$\eta_{2c}$ for the CAF 
phases. Our spin-wave expansion for the CAF phase becomes unreliable for 
$\eta>0.95$. We extrapolate our data to find $\eta_{2c}=0.58$ for $\zeta=1$. The 
dotted line is the extrapolated curve. The dashed line in the middle represents
the classical first-order phase transition line $\eta_c^{\rm class}=\zeta/2$. 
The spin-gap ($\eta_{2c}-\eta_{1c}$) increases with the anisotropy parameter $\zeta$. (color online)}
\end{figure}
Phase diagram for the $J_1-J_1^\prime-J_2$ model is displayed in Fig.~\ref{fig:critical}. 
The solid lines indicate the critical points $\eta_{1c}$ for the AF and 
$\eta_{2c}$ for the CAF 
phases. The dashed line is the classical first-order phase transition line between the 
two phases.  Our spin-wave expansion for the CAF phase becomes unreliable for 
$\eta>0.95$. Thus we extrapolate our data to obtain $\eta_{2c}=0.58$ for $\zeta=1$. The 
dotted line in Fig.~\ref{fig:critical} is the extrapolated curve. This result $\eta_{2c}=0.58$ is
in good agreement with the expected value of $\approx 0.60$ for the isotropic case. 
Figure~\ref{fig:critical} shows that the spin-gap ($\eta_{2c}-\eta_{1c}$) increases with 
increase in $\zeta$.

Within our
spin-wave expansion we do not find any quantum triple point for any values of $\zeta$ and $\eta$. This is 
in contrary to the findings in Refs.~\onlinecite{viana07}, ~\onlinecite{mendonca10}, and \onlinecite{bishop08}. Instead 
from our calculations we find that there are two ordered phases
separated by the magnetically disordered phase. Our proposed 
phase diagram is consistent with the phase diagram obtained by the 
DMRG calculations~\cite{hako01}, 
exact diagonalization method~\cite{sindzingre04}, and 
the results from the continuum limit of the present model~\cite{oleg04}. Our results are also 
in accord with the numerical evidence of a dimerized intermediate phase in the frustrated two leg model
up to a certain value of the interchain coupling.~\cite{liu08,tosh10}

\section{\label{sec:conclusions}Conclusions}
In this work for an antiferromagnetic square lattice we have provided a comprehensive study of the effects of quantum fluctuations due to spatial anisotropy and frustration between nearest and next-nearest neighbors on the low-temperature thermodynamic properties of the two ordered phases of the system. 
Using second-order spin-wave expansion we have calculated the spin wave energy in the entire Brillouin zone, renormalized spin-wave velocities, and the magnetizations for the antiferromagnetic Ne\'{e}l and columnar antiferromagnetic phases. 
We have found that the second-order corrections contribute significantly to stabilize the quantum phase diagram of the system as frustration between the spins increase. As expected from linear spin wave theory magnetization becomes zero at the quantum critical points. However, the second-order corrections slightly extend the region of the AF-order. 

Our results for the spin-wave energies are compared with the
recent experimental results using neutron scattering for CFTD.~\cite{chris07} With our second-order 
spin-wave expansion we have reproduced the previous numerical results that the spin-wave energy at 
$(\pi,0)$ is about 1.4\% smaller that at $(\pi/2,\pi/2)$. This result falls short of the experimental result. Furthermore, we find that the dip in spin wave energy at $(\pi,0)$ increases with increase in NNN frustration. This can provide a measure of the effect of
frustration experimentally. For a few values of small frustration and anisotropy we have explicitly calculated the percentage changes between the spin-wave energies at 
$(\pi,0)$ and at $(\pi/2,\pi/2)$. We have shown how the renormalized spin-wave velocities along the row
and column direction change with frustration. Both these velocities become zero close to the critical
transition points. For the AF-ordered phase we have also calculated the spin deviation from the 
classical value of 0.5  with no NNN coupling for different values of directional anisotropies and have
obtained an empirical equation based on our numerical data. 


For the CAF-ordered phase we have obtained similar results. The magnetization becomes zero at the quantum critical points as frustration increases. For $\zeta<0.95$ our
calculations produce correct results but our present numerical approach is not reliable for $\zeta>0.95$. 
Thus we were not able to find the thermodynamic properties for $\zeta=1$.  Based on our data we have extrapolated the magnetization 
for the  case $\zeta=1, \eta=1$ and found it to be 0.30 which is in good agreement with existing experimental result~\cite{melzi00,melzi01,carretta02,bombardi04}. Our extrapolated value
of the quantum critical point $\eta_{2c}=0.58$ for $\zeta=1, \eta=1$ is also in good agreement
with the expected value $\approx 0.60$. We have not found much experimental data on this system to compare with our other results such as the spin-wave energy dispersion in the entire BZ and
the spin-wave velocities. 

Finally we combined our results for the magnetization of the two phases with different directional 
anisotropies to obtain the complete magnetic phase diagram of the system. We have found 
that two ordered phases are always
separated by the disordered paramagnetic phase. Our proposed phase diagram is consistent with the phase diagram obtained from the DMRG calculations~\cite{hako01}, exact diagonalization method~\cite{sindzingre04}, and the results from the continuum limit of the present model~\cite{oleg04}.
Our results are also in accord with the numerical evidence of a dimerized intermediate phase in the frustrated two leg model up to a certain value of the interchain coupling.  In summary with our present approach based on second-order
spin-wave expansion we do not find existence of quantum triple points for any values of $\zeta$ and $\eta$.

\section{Acknowledgment}
The author would like to thank M. Kr\'{c}mar, C. J. Hammer, A. L. Chernyshev for useful discussions and comments and A. Genz, N. Woody for computational help. This project acknowledges the use of
the Cornell Center for Advanced Computing's ``MATLAB on the TeraGrid'' experimental
computing resource funded by NSF grant 0844032 in partnership with Purdue 
University, Dell, The MathWorks, and Microsoft.
\appendix
\section{\label{VertexAF} Vertex factors for the AF phase}
The six vertex factors for the AF-phase are given below. 
\bea
V_{1234}^{(1)} &=& \gamma_1 (1-4)x_1x_4+\gamma_1(1-3)x_1x_3+\gamma_1(2-4)x_2x_4
+\gamma_1(2-3)x_2x_3 \non \\
&-& \half \Big[\gamma_1(1)x_1 +\gamma_1 (2) x_2 +\gamma_1(3) x_3+\gamma_1(4)x_4
+\gamma_1 (2-3-4)x_2 x_3 x_4 \non \\
&+& \gamma_1(1-3-4)x_1x_3x_4 + \gamma_1(4-2-1)x_1x_2x_4+\gamma_1(3-2-1)x_1x_2x_3\Big]\non \\
&-& \Big(\frac {2\eta}{1+\zeta}\Big) f_{1234}\Big[1+{\rm sgn}(\gamma_{\bf G})x_1x_2x_3x_4
\Big],\\
V_{1234}^{(2)} &=& \gamma_1 (2-4)x_1x_3+\gamma_1(1-4)x_2x_3+\gamma_1(2-3)x_1x_4
+\gamma_1(1-3)x_2x_4 \non \\
&-& \half \Big[\gamma_1(2)x_1x_3x_4 +\gamma_1 (1) x_2x_3x_4 +\gamma_1(4) x_1x_2x_3+
\gamma_1(3)x_1x_2x_4 \non \\
&+&\gamma_1 (2-3-4)x_1
+ \gamma_1(1-3-4)x_2 + \gamma_1(4-2-1)x_3+\gamma_1(3-2-1)x_4\Big]\non \\
&-& \Big(\frac {2\eta}{1+\zeta}\Big) f_{1234}\Big[x_1x_2x_3x_4+{\rm sgn}(\gamma_{\bf G})
\Big],\\
V_{1234}^{(3)} &=&\gamma_1 (2-4)+\gamma_1(1-3)x_1x_2x_3x_4+\gamma_1(1-4)x_1x_2
+ \gamma_1(2-3)x_3x_4  \non \\
&-& \half \Big[\gamma_1(2)x_4 +\gamma_1 (1) x_1x_2x_4 +\gamma_1(2-3-4) x_3+
\gamma_1(1-3-4)x_1x_2x_3 \non \\
&+&\gamma_1 (4)x_2
+ \gamma_1(3)x_2 x_3x_4+ \gamma_1(4-2-1)x_1+\gamma_1(3-2-1)x_1x_3x_4\Big]\non \\
&-& \Big(\frac {2\eta}{1+\zeta}\Big) f_{1234}\Big[x_2x_4+{\rm sgn}(\gamma_{\bf G})x_1x_3\Big],\\
V_{1234}^{(4)} &=&-\gamma_1 (2-4)x_4-\gamma_1(1-4)x_1x_2x_4-\gamma_1(2-3)x_3
- \gamma_1(1-3)x_1x_2x_3  \non \\
&+& \half \Big[\gamma_1(2) +\gamma_1 (1) x_1x_2 +\gamma_1(3) x_2x_3+
\gamma_1(4)x_2x_4 \non \\
&+&\gamma_1 (2-3-4)x_3x_4
+ \gamma_1(1-3-4)x_1x_2 x_3x_4+ \gamma_1(3-2-1)x_1x_3+\gamma_1(4-2-1)x_1x_4\Big]\non \\
&+& \Big(\frac {2\eta}{1+\zeta}\Big) f_{1234}\Big[x_2+{\rm sgn}(\gamma_{\bf G})x_1x_3x_4
\Big],\\
V_{1234}^{(5)} &=&-\gamma_1 (2-4)x_1-\gamma_1(2-3)x_1x_3x_4-\gamma_1(1-4)x_2
- \gamma1018_1(1-3)x_2x_3x_4  \non \\
&+& \half \Big[\gamma_1(2)x_1x_4 +\gamma_1 (1) x_2x_4 +\gamma_1(4) x_1x_2+
\gamma_1(3)x_1x_2x_3x_4 \non \\
&+&\gamma_1 (2-3-4)x_1x_3
+ \gamma_1(1-3-4)x_2 x_3+ \gamma_1(4-2-1)+\gamma_1(3-2-1)x_3x_4\Big]\non \\
&+& \Big(\frac {2\eta}{1+\zeta}\Big) f_{1234}\Big[x_1x_2x_4+{\rm sgn}(\gamma_{\bf G})x_3
\Big],\\
V_{1234}^{(6)} &=&\gamma_1 (2-4)x_2x_3+\gamma_1(2-3)x_2x_4+\gamma_1(1-3)x_1x_4
+ \gamma_1(1-4)x_1x_3  \non \\
&-& \half \Big[\gamma_1(2)x_2x_3x_4 +\gamma_1 (3)x_4 +\gamma_1(2-3-4)x_2+
\gamma_1(3-2-1)x_1x_2x_4 \non \\
&+&\gamma_1 (1)x_1x_3x_4
+ \gamma_1(4) x_3+ \gamma_1(1-3-4)x_1+\gamma_1(4-2-1)x_1x_2x_3\Big]\non \\
&-& \Big(\frac {2\eta}{1+\zeta}\Big) f_{1234}\Big[x_3x_4+{\rm sgn}(\gamma_{\bf G})x_1x_2
\Big],
\eea
with
\be
f_{1234}=\half\Big[\gamma_2(1-3)+\gamma_2(1-4)+\gamma_2(2-3)+\gamma_2(2-4)
-\gamma_2(1)-\gamma_2(2)-\gamma_2(3)-\gamma_2(4) \Big].
\ee

\section{\label{Greensfunction} Green's function and Magnon Self energy for the AF phase}
\begin{figure}[httb]
\centering
\includegraphics[width=4.5in]{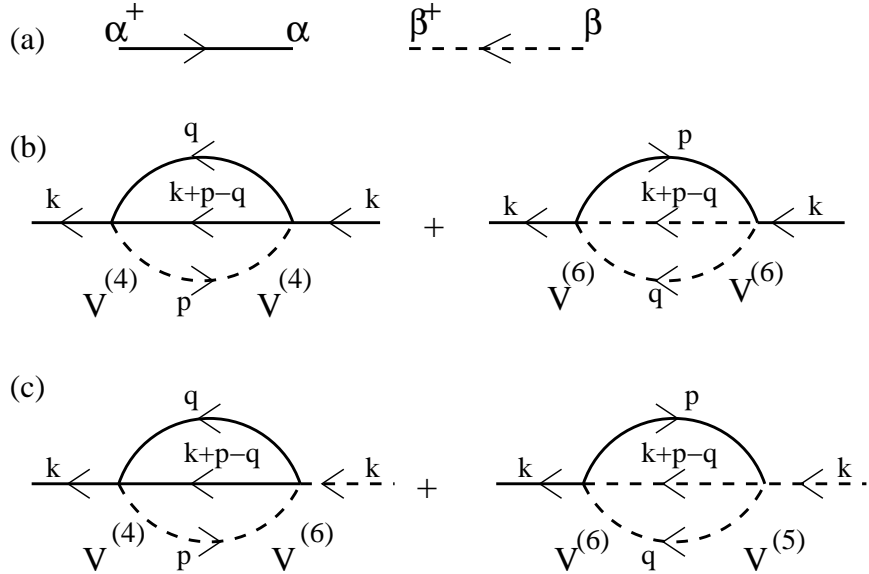}
\caption{\label{fig:Feyn} (a) The solid and the dashed lines correspond to the
$\alpha$ and $\beta$ propagators. Second-order diagrams for the self-energies 
$\Sigma_{\alpha \alpha}({\bf k},\omega)$ and $\Sigma_{\alpha \beta}({\bf k},\omega)$
are shown in (b) and (c).
$V^{(4)}, V^{(5)}, V^{(6)}$ are the vertex factors (see text).}
\end{figure}

The time-ordered magnon Green's functions are defined as
\begin{eqnarray*}
G_{\alpha \alpha} ({\bf k},t) &=& -i\langle T(\alpha_{\bf k}(t)\alpha^\dag_{\bf k}(0))\rangle,\;\;\;
G_{\beta \beta} ({\bf k},t) = -i\langle T(\beta^\dag_{-\bf k}(t)\beta_{-\bf k}(0))\rangle,\\
G_{\alpha \beta} ({\bf k},t) &=& -i\langle T(\alpha_{\bf k}(t)\beta_{-\bf k}(0))\rangle,\; \;\;
G_{\beta \alpha} ({\bf k},t) = -i\langle T(\beta^\dag_{-\bf k}(t)\alpha^\dag_{\bf k}(0))\rangle, 
\end{eqnarray*}
Considering $H_0$ as the unperturbed Hamiltonian the Fourier transformed 
unperturbed propagators are given as
\bea
G^0_{\alpha \alpha} ({\bf k},\omega) &=& [\omega - E_k+i\delta]^{-1},\;\;\;
G^0_{\beta \beta} ({\bf k},\omega) = [-\omega - E_k+i\delta]^{-1}, \\
G^0_{\alpha \beta} ({\bf k},\omega) &=& G^0_{\beta \alpha}({\bf k},\omega)=0,
\eea
with $\delta \rightarrow 0+$. The magnon energy 
$E_{\bf k}=\kappa_{\bf k}\epsilon_{\bf k}$ is measured in units of $J_1Sz(1+\zeta)$.
The graphical representation of the Green functions are shown in Fig.~\ref{fig:Feyn}(a).
The full propagators $G_{ij}({\bf k},\omega)$ satisfy the matrix Dyson equation:
\be
G_{ij}({\bf k},\omega)=G^{0}_{ij}({\bf k},\omega)+\sum_{mn}G^0_{im}({\bf k},\omega)
\Sigma_{mn}({\bf k},\omega)G_{nj}({\bf k},\omega),
\ee
where the self-energy $\Sigma_{ij}({\bf k}$ can be expressed in
powers of $1/(2S)$ as 
\be
\Sigma_{ij}({\bf k},\omega)=\frac 1{(2S)}\Sigma_{ij}^{(1)}({\bf k},\omega)
+\frac 1{(2S)^2}\Sigma_{ij}^{(2)}({\bf k},\omega) + ....
\ee
Using the relations
\[
V^{(5)}_{[{\bf k+p-q}],{\bf q,p,k}}={\rm sgn}(\gamma_{\bf G})V^{(4)}_{{\bf k,p,q},[{\bf k+p-q}]},\;\;
V^{(6)}_{{\bf q}, [{\bf k+p-q}],{\bf k,p}}={\rm sgn}(\gamma_{\bf G})V^{(6)}_{{\bf k,p,q},[{\bf k+p-q}]},
\]
the first and second order self-energies are written as
\bea
\Sigma_{\alpha \alpha}^{(1)}({\bf k},\omega) &=& \Sigma_{\beta \beta}^{(1)}({\bf k},\omega)=A_{\bf k}, \\
\Sigma_{\alpha \beta}^{(1)}({\bf k},\omega) &=& \Sigma_{\beta \alpha}^{(1)}({\bf k},\omega)=B_{\bf k}, \\
\Sigma_{\alpha \alpha}^{(2)}({\bf k},\omega) &=& \Sigma_{\beta \beta}^{(2)}(-{\bf k},-\omega)
=C_{1{\bf k}} 
+ \Big(\frac {2}{N} \Big)^2\sum_{{\bf pq}}2l_{\bf k}^2l_{\bf p}^2l_{\bf q}^2l_{\bf k+p-q}^2 \non \\
&\times& \Big[\frac {|V^{(4)}_{\bf k,p,q,[k+p-q]}|^2}{\omega -E_{\bf p}
-E_{\bf q}-E_{\bf k+p-q}+i\delta} - \frac {|V^{(6)}_{\bf k,p,q,[k+p-q]}|^2}{\omega +E_{\bf p}
+E_{\bf q}+E_{\bf k+p-q}-i\delta}  \Big],\label{sigma1}\\
\Sigma_{\alpha \beta}^{(2)}({\bf k},\omega) &=& \Sigma_{\beta \alpha}^{(2)}(-{\bf k},-\omega)
=C_{2{\bf k}} 
+ \Big(\frac {2}{N} \Big)^2\sum_{{\bf pq}}2l_{\bf k}^2l_{\bf p}^2l_{\bf q}^2l_{\bf k+p-q}^2 {\rm sgn}(\gamma_{\bf G})\non \\
&\times& V^{(4)}_{\bf k,p,q,[k+p-q]}V^{(6)}_{\bf k,p,q,[k+p-q]}
\frac{2(E_{\bf p}+E_{\bf q}+E_{\bf k+p-q})}{\omega^2-(E_{\bf p}+E_{\bf q}+E_{\bf k+p-q})^2},
\label{sigma2}
\eea
where $[{\bf k+p-q}]$ is mapped to $({\bf k+p-q})$ in the first BZ by the reciprocal 
vector ${\bf G}$. Feynman diagrams for the second-order self energies are shown 
in Fig.~\ref{fig:Feyn}(b)-(c). Above the coefficients
$C_{1{\bf k}}$ and $C_{2{\bf k}}$ are 
\bea
C_{1{\bf k}} &=& \half l_k^2 \Big(\frac {2}{N}\Big)^2 \sum_{12} l_1^2l_2^2\Big[
-6\gamma_1(2-1-{\bf k})x_{\bf k}x_1x_2 + \gamma_1(2) x_1^2x_2+
\gamma_1(2)x_{\bf k}^2x_1^2x_2 \non \\
&+& 2\gamma_1({\bf k})x_{\bf k}x_1^2 
+\gamma_1(1)x_{\bf k}^2x_1+\gamma_1(2)x_2\Big]
- \frac 1{4}\Big(\frac {2\eta}{1+\zeta} \Big) l_{\bf k}^2(1+x_{\bf k}^2)
{\tilde C}_{\bf k}, \\
C_{2{\bf k}} &=& \half l_k^2 \Big(\frac {2}{N}\Big)^2 \sum_{12} l_1^2l_2^2\Big[
3\gamma_1(2-1-{\bf k})x_1x_2 + 3\gamma_1(2-1-{\bf k}) x_{\bf k}^2x_1x_2-
2\gamma_1(1)x_{\bf k}x_1x_2^2 \non \\
&-& 2\gamma_1(2)x_{\bf k}x_2 
-\gamma_1({\bf k})x_2^2-\gamma_1({\bf k})x_{\bf k}^2x_2^2\Big]
- \frac 1{2}\Big(\frac {2\eta}{1+\zeta} \Big) l_{\bf k}m_{\bf k}
{\tilde C}_{\bf k},
\label{C1C2}
\eea
with
\bea
{\tilde C}_{\bf k} &=& \Big(\frac {2}{N}\Big)^2 \sum_{12} l_1^2l_2^2\Big\{
\Big[2\gamma_2({\bf k})+\gamma_2(1)+\gamma_2(2)-4\gamma_2({\bf k}+1-2)\Big]x_1^2 \non \\
&+& \Big[\gamma_2(2)-\gamma_2(1+2-{\bf k})\Big](1+x_1^2x_2^2) \Big\}.
\label{Ceqn}
\eea
The divergent terms in $C_{1{\bf k}}$ and 
$C_{2{\bf k}}$  for ${\bf k} \rightarrow 0$ are canceled out by the
second terms in Eqs.~\ref{sigma1} and \ref{sigma2}.~\cite{igar92}  

\section{\label{VertexCAF} Second order Coefficients and Vertex factors for the CAF phase}
The coefficients that appear in the second-order corrections in the Hamiltonian for the CAF-phase
are:
\bea
C_{1{\bf k}}^\prime &=& \half l_k^{\prime 2} \Big(\frac {2}{N}\Big)^2 \sum_{12} l_1^{\prime 2}l_2^{\prime 2}\Big[
-6\gamma_1^\prime(2-1-{\bf k})x^\prime_{\bf k}x^\prime_1 x^\prime_2 + \gamma^\prime_1(2) x^{\prime 2}_1x^\prime_2+
\gamma^\prime_1(2)x_{\bf k}^{\prime 2}x_1^{\prime 2}x^\prime_2 \non \\
&+& 2\gamma^\prime_1({\bf k})x^\prime_{\bf k}x_1^{\prime 2} 
+\gamma^\prime_1(1)x_{\bf k}^{\prime 2}x^\prime_1+\gamma^\prime_1(2)x^\prime_2\Big]
- \frac 1{4}\Big(\frac {\zeta}{1+2\eta} \Big) l_{\bf k}^{\prime 2}(1+x_{\bf k}^{\prime 2})
{\tilde C}_{\bf k}^\prime, \\
C_{2{\bf k}}^\prime &=& \half l_k^{\prime 2} \Big(\frac {2}{N}\Big)^2 \sum_{12} l_1^{\prime 2}l_2^{\prime 2}\Big[
3\gamma^\prime_1(2-1-{\bf k})x^\prime_1x^\prime_2 + 3\gamma^\prime_1(2-1-{\bf k}) x_{\bf k}^{\prime 2}x^\prime_1x^\prime_2-
2\gamma^\prime_1(1)x^\prime_{\bf k}x^\prime_1x_2^{\prime 2} \non \\
&-& 2\gamma^\prime_1(2)x^\prime_{\bf k}x_2^\prime 
-\gamma^\prime_1({\bf k})x_2^{\prime 2}-\gamma^\prime_1({\bf k})x_{\bf k}^{\prime 2}x_2^{\prime 2}\Big]
- \frac 1{2}\Big(\frac {\zeta}{1+2\eta} \Big) l_{\bf k}^\prime m_{\bf k}^\prime
{\tilde C}_{\bf k}^\prime,
\eea
where
\be
l_{\bf k}^\prime = \Big[\frac {1+\epsilon^\prime_{\bf k}}{2\epsilon^\prime_{\bf k}} \Big]^{1/2},\;\;
m_{\bf k}^\prime = -{\rm sgn}(\gamma^\prime_{\bf k})\Big[\frac {1-\epsilon^\prime_{\bf k}}
{2\epsilon^\prime_{\bf k}} \Big]^{1/2}=-x_{\bf k}^\prime l_{\bf k}^\prime, 
\ee
and 
\bea
{\tilde C}_{\bf k}^\prime &=& \Big(\frac {2}{N}\Big)^2 \sum_{12} l_1^{\prime 2}l_2^{\prime 2}\Big\{
\Big[2\gamma^\prime_2({\bf k})+\gamma^\prime_2(1)+\gamma^\prime_2(2)-4\gamma^\prime_2({\bf k}+1-2)\Big]x_1^{\prime 2} \non \\
&+& \Big[\gamma^\prime_2(2)-\gamma^\prime_2(1+2-{\bf k})\Big](1+x_1^{\prime 2}x_2^{\prime 2}) \Big\}.
\label{Ceqn2}
\eea
For the magnetization and spin-wave dispersion calculations only the vertex 
factors $V^{\prime (4)}$ and $V^{\prime (6)}$ are required
which are:
\bea
V_{1234}^{\prime (4)} &=&-\gamma^\prime_1 (2-4)x^\prime_4-\gamma^\prime_1(1-4)x^\prime_1 x^\prime_2x^\prime_4-
\gamma^\prime_1(2-3)x^\prime_3
- \gamma^\prime_1(1-3)x^\prime_1x^\prime_2x^\prime_3  \non \\
&+& \half \Big[\gamma^\prime_1(2) +\gamma^\prime_1 (1) x^\prime_1x^\prime_2 +\gamma^\prime_1(3) x^\prime_2x^\prime_3+
\gamma^\prime_1(4)x^\prime_2x^\prime_4 \non \\
&+&\gamma^\prime_1 (2-3-4)x^\prime_3x^\prime_4
+ \gamma^\prime_1(1-3-4)x^\prime_1x^\prime_2 x^\prime_3x^\prime_4+ 
\gamma^\prime_1(3-2-1)x^\prime_1x^\prime_3+\gamma^\prime_1(4-2-1)x^\prime_1x^\prime_4\Big]\non \\
&+& \Big(\frac {\zeta}{1+2\eta}\Big) f^\prime_{1234}\Big[x^\prime_2+{\rm sgn}(\gamma^\prime_{\bf G})x^\prime_1x^\prime_3x^\prime_4
\Big],\\
V_{1234}^{\prime (6)} &=&\gamma^\prime_1 (2-4)x^\prime_2x^\prime_3+
\gamma^\prime_1(2-3)x^\prime_2x^\prime_4+\gamma^\prime_1(1-3)x^\prime_1x^\prime_4
+ \gamma^\prime_1(1-4)x^\prime_1x^\prime_3  \non \\
&-& \half \Big[\gamma^\prime_1(2)x^\prime_2x^\prime_3x^\prime_4 +
\gamma^\prime_1 (3)x^\prime_4 +\gamma^\prime_1(2-3-4)x^\prime_2+
\gamma^\prime_1(3-2-1)x^\prime_1x^\prime_2x^\prime_4 \non \\
&+&\gamma^\prime_1 (1)x^\prime_1x^\prime_3x^\prime_4
+ \gamma^\prime_1(4) x^\prime_3+ \gamma^\prime_1(1-3-4)x^\prime_1+
\gamma^\prime_1(4-2-1)x^\prime_1x^\prime_2x^\prime_3\Big]\non \\
&-& \Big(\frac {\zeta}{1+2\eta}\Big) f^\prime_{1234}\Big[x^\prime_3x^\prime_4+
{\rm sgn}(\gamma^\prime_{\bf G})x^\prime_1x^\prime_2
\Big],
\eea
with
\be
f^\prime_{1234}=\half\Big[\gamma^\prime_2(1-3)+\gamma^\prime_2(1-4)+\gamma^\prime_2(2-3)+\gamma^\prime_2(2-4)
-\gamma^\prime_2(1)-\gamma^\prime_2(2)-\gamma^\prime_2(3)-\gamma^\prime_2(4) \Big].
\ee
\bibliography{Aniso2D}

\end{document}